\documentclass[aps,prb,notitlepage,superscriptaddress,showpacs]{revtex4-1}
\usepackage{graphicx,subfigure,epsfig}
\usepackage{dcolumn}
\usepackage{amssymb}
\usepackage{times}
\usepackage{amsmath}
\usepackage{amsfonts}
\usepackage{mathrsfs}
\usepackage{setspace}
\usepackage{latexsym}
\usepackage{bbm}
\usepackage{float}
\usepackage{flafter}
\usepackage{bm}
\usepackage{epstopdf}
\usepackage{hyperref}
\usepackage{color}
\usepackage{multirow}
\newcommand{\I}{\mathrm{i}}
\newcommand{\E}{\mathrm{e}}

\begin{document}

\title{Exact sign structure of the $t$-$J$ chain and the single hole ground state}
\author{Zheng Zhu}
\affiliation{Institute for Advanced Study, Tsinghua University, Beijing, 100084, China}
\author{Qing-Rui Wang}
\affiliation{Institute for Advanced Study, Tsinghua University, Beijing, 100084, China}
\author{D. N. Sheng}
\affiliation{Department of Physics and Astronomy, California State University, Northridge, CA, 91330, USA}
\author{Zheng-Yu Weng}
\affiliation{Institute for Advanced Study, Tsinghua University, Beijing, 100084, China}
\affiliation{Collaborative Innovation Center of Quantum Matter, Tsinghua University, Beijing, 100084, China}

\date{\today}


\begin{abstract}
Injecting a single hole into a one-dimensional Heisenberg spin chain is probably the simplest case of doping a Mott insulator. The motion of such a single hole will generally induce a many-body phase shift, which can be identified by an exact sign structure of the model known as the phase string. We show that the sign structure is nontrivial even in this simplest problem, which is responsible for the essential properties of Mott physics. We find that the characteristic momentum structure, the Luttinger liquid behavior, and the quantum phase interference of the hole under a periodic boundary condition, can all be attributed to it. We use the density matrix renormalization group (DMRG) numerical simulation to make a comparative study of the $t$-$J$ chain and a model in which the sign structure is switched off. We further show that the key DMRG results can be reproduced by a variational wave function with incorporating the correct sign structure. Physical implications of the sign structure for doped Mott insulators in general are also discussed.
\end{abstract}

\maketitle

\section{Introduction}
\label{sec:intro}

How a doped hole propagates in a ``vacuum'' that is full of quantum spins is a central question in a doped Mott system \cite{Anderson,Lee2006}. On general grounds, one expects a ``cloud of spin excitations'' to be generated to accompany the motion of the hole. In a more conventional/ weakly correlated system, a similar ``cloud'' forming around a testing particle is usually finite in size, remains featureless and rigid at low energy, which dresses only the particle's effective mass.  The challenge arises, however, if the spin cloud becomes neither featureless nor rigid, or in other words, the motion of the hole becomes strongly-correlated in nature. Generally speaking, a new mathematical description will be needed here. The issue of the single hole problem has attracted intense attention since the discovery of the high-$T_c$ cuprate, which is considered to be a doped Mott insulator \cite{Anderson1987}.

In literature, the one-dimensional (1D)  doped Mott systems have been well studied. The exact Bethe ansatz or the Lieb-Wu solution \cite{Lieb1968} exists for the Hubbard model at an arbitrary doping concentration and ratio of the on-site Coulomb repulsion $U$ and the nearest-neighbor hopping integral $t$. In particular, the 1D $t$-$J$ model can be solved exactly at $t/J\rightarrow \infty$ \cite{Woynarovich1982,Ogata1990,Ogata1991b} and $t/J=1/2$ \cite{Sutherland1975,Schlottmann1987,Bares1990,Bares1991} ($J$ is the superexchange coupling), both of which behave like a Luttinger liquid \cite{Haldane1981a,Haldane1981b} at finite doping. Numerically, the phase diagram has been also given via exact diagonalization (ED) \cite{Ogata1991,Dagotto1994} and the density matrix renormalization group (DMRG) \cite{Moreno2011} methods. As a matter of fact, based on the 1D exact solution, Anderson proposed \cite{Anderson90PRL} the idea of the unrenormalizable many-body phase shift, which is argued \cite{Anderson,Ren-Anderson1993} to be generally responsible for the Luttinger liquid behavior in the doped Mott insulator. The quantitative characterization of such phase shift was later analytically identified \cite{Weng1991,Weng1992,Wang2014} in the 1D $t$-$J$ model. The ground state properties of the doped Hubbard model at $U\gg t$ or the $t$-$J$ model at $J\ll t$ can be also approximated by the so-called squeezed spin chain description\cite{Ogata1990,Ogata1991b,Weng1991,Weng1992,Shiba1992,Weng1994,Weng1997,Kruis2004}.

Nevertheless, a simple microscopic understanding is still much needed, even for the simplest one-hole-doped 1D case. By answering the question raised at the beginning of this paper, one may gain a deeper insight into the strong correlation nature of the Mott physics, which goes beyond the specific 1D geometry. Utilizing exact analysis and numerical methods, one hopes to clearly illustrates the single-hole's motion in an antiferromagnetic spin background qualitatively and quantitatively, which are relatively easier to handle in 1D. It may then provide important insights for the problem in two dimensions (2D), which is more relevant to cuprate superconductors and other strongly correlated materials.

In this paper, we investigate the ground state of the one-hole-doped 1D Heisenberg chain using the exact analysis, DMRG, and wave function approach based on a variational Monte Carlo (VMC) method. The main results are obtained as follows. First of all, we explicitly show that a nontrivial sign structure or phase string emerges once a hole is doped into the Heisenberg spin chain, which otherwise is statistical-sign free. Since such sign structure is present for any dimensions, the 1D limit provides the simplest example to show its novel consequences. The detailed analyses will be given in Sec. II; Secondly, in contrast to a bare hole state created by annihilating an electron in the half-filled ground state, the true hole ground state differs by a fundamentally changed momentum distribution, as well as the vanishing single-particle spectral weight, obeying a power-law scaling with the length of the chain. The phase string induced by one hole doping, as the singular many-body phase shift contributed by the spins in the vacuum, is responsible for the above momentum readjustment and the Luttinger liquid behavior. These will be confirmed by the DMRG simulations in Sec. III; Furthermore, we show that a variational wave function constructed by incorporating the correct sign structure or the phase string, can reproduce the DMRG results by using the VMC calculation in Sec. IV; Finally, in the summary section (Sec. V), we will also discuss how the phase string sign structure plays a critical role in a general doped Mott insulator beyond the 1D case examined in the present work.

\section{The model and exact analysis}
\label{sec:model}

\subsection{The model}

In this paper,  the main focus will be the ground state properties of a single hole injected into a 1D Mott insulator. It is described by the $t$-$J$ Hamiltonian, $H_{t\text{-}J} = H_t+H_J$, generally defined in the Hilbert space constrained by the no-double-occupancy condition as follows
\begin{equation}
\begin{split}
H_t &= -t \sum_{\langle {ij}\rangle, \sigma } {({c_{i\sigma }^{\dag}c_{j\sigma }+\mathrm{h.c.}})}, \\
H_J &= J \sum_{\langle {ij}\rangle } {\left(\mathbf{S}_{i}\cdot \mathbf{S}_{j}-\frac{1}{4}n_{i}n_{j}\right)}.
\end{split}
\label{Eq:TJModel}
\end{equation}
Here, the operator ${c_{i\sigma }^{\dagger }}$ creates an electron at site $i$ with spin $\sigma $, and ${\mathbf{S}_{i}}$ is the spin operator, ${n_{i}}$ the number
operator, respectively, with the summations running over all the nearest-neighbors $\langle ij\rangle $ along the chain.

For a comparative study, we will also consider the ground state of the so-called $\sigma$$\cdot$$t$-$J$ model \cite{ZZ2013}, $H_{\sigma\cdot t\text{-}J} = H_{\sigma\cdot t} +H_J$, in which the superexchange term $H_J$ remains the same as in the $t$-$J$ model Eq.~(\ref{Eq:TJModel}), but the hopping term is modified by
\begin{equation}
H_{\sigma \cdot t} = -t \sum_{\langle {ij}\rangle, \sigma }\sigma {({c_{i\sigma}^{\dag }c_{j\sigma }+\mathrm{h.c.}})},
\label{Eq:SigmaTJModel}
\end{equation}
in which an extra spin-dependent sign $\sigma=\pm$ is inserted. The distinction between the ground states of the $t$-$J$ and the $\sigma$$\cdot$$t$-$J$ models will reveal the critical role of the phase-string sign structure hidden in the $t$-$J$ Hamiltonian, which however is precisely eliminated by the sign $\sigma$ in $H_{\sigma \cdot t}$ in Eq.~(\ref{Eq:SigmaTJModel}), as to be seen in the following \cite{ZZ2013}.

Finally, these two models may be connected by tuning the spin-dependent hopping integrals as follows
\begin{eqnarray}
H_{t_\uparrow \text{-}t_\downarrow} &=& - t_\uparrow\sum_{\langle {ij}\rangle}
{c_{i\uparrow}^{\dag }c_{j\uparrow }} - t_\downarrow\sum_{\langle {ij}\rangle}
{c_{i\downarrow}^{\dag }c_{j\downarrow }}+\mathrm{h.c.}
\label{Eq:tutdJmodel}
\end{eqnarray}
Here, $t_\uparrow$ and $t_\downarrow$ are the hopping amplitudes for the up-spin and down-spin electrons, respectively, and again the constraint $n_{i}\leq 1$ in the Hilbert space is always enforced. When $t_\uparrow=t_\downarrow=t$, the above model becomes the normal $t$-$J$ model in Eq.~(\ref{Eq:TJModel}). Similarly, when $t_\uparrow=t$ and $t_\downarrow=-t$, this model becomes the $\sigma$$\cdot$$t$-$J$ model in Eq.~(\ref{Eq:SigmaTJModel}). Then by fixing $t_\uparrow=t$ and tuning the hopping integral $t_\downarrow$, one may continuously connect these two models to turn on or off the phase string sign structure. Note that the spin rotational symmetry is slightly broken here in the x-y plane by the hopping term involving a spin-1/2 in the one-hole-doping case, while the background spins governed by $H_J$ still obey the spin rotational symmetry.

\subsection{The exact sign structure}
\label{sec:exactsign}

For one-hole-doped $t$-$J$ model Eq.~(\ref{Eq:TJModel}) on a bipartite lattice, it has been previously demonstrated that the hopping of the hole will pick up a sequence of signs, i.e., $(+1)\times (-1)\times (-1)\times \cdots$,  known as a phase string \cite{Sheng1996,Weng1997,Wu2008sign}.  Here, the sign $\pm $ keeps track of the microscopic process of an $\uparrow$ or $\downarrow$-spin exchanging with the hole at each step of hopping. The exact sign structure of the $t$-$J$ model with one hole is precisely given by
\begin{equation}\label{tauc}
\tau _c  = \left( { - 1} \right)^{N_h^ \downarrow  [c]} ~,
\end{equation}
with $N_{h}^{\downarrow }[c]$ denoting the total number of exchanges between the hole and down spins along a path $c$ which can be either open or closed.

For example, $\tau_c$ appears in the single-particle propagator of the hole from $i$ to $j$ as follows \cite{Sheng1996,Weng1997}
\begin{equation}
G_{1h}(j,i;E)\propto \sum_{c}{\tau }_{c}\mathcal{W}[c]~,  \label{gtj}
\end{equation}
where $\{c\}$ includes all the spin and hole paths with the path weight $\mathcal{W}[c]\ge 0$ at energy $E<0$. Among $\{c\}$, the path of the hole, which connects $i$ and $j$, is an \emph{open} one \cite{Sheng1996,Weng1997}.

On the other hand, $\tau_c$ also appears in the partition function \cite{Wu2008sign}
\begin{equation}
Z_{t\text{-}J}=\sum_{c}{\tau }_{c}\mathcal{Z}[c]~,  \label{ztj}
\end{equation}
where $\mathcal{Z}[c]\ge 0$ for any \emph{closed} path $c$, which is generally temperature- as well as $t$- and $J$-dependent.

Furthermore, such phase-string sign structure can be artificially switched off by introducing the $\sigma$$\cdot$$t$-$J$ model Eq.~(\ref{Eq:SigmaTJModel}) with inserting a spin-dependent sign  $\sigma $ in the hopping term.  It is easy to show \cite{ZZ2013} that the phase string disappears in the $\sigma$$\cdot$$t$-$J$ model, with $\tau _c $ replaced by $+1 $ in, say, Eq.~(\ref{ztj})
\begin{equation}
Z_{\sigma\cdot t\text{-}J}=\sum_{c}\mathcal{Z}[c]~,  \label{zstj}
\end{equation}
where the non-negative weight $\mathcal{Z}[c]$ for each path $c$ remains unchanged. Similarly $\tau _c $ is also precisely eliminated in Eq.~(\ref{gtj}). Therefore, the $t$-$J$ and the $\sigma$$\cdot$$t$-$J$ models are solely differentiated by the presence and absence of the phase string sign structure $\tau _c$. In other words, the distinction between the two ground states will uniquely reveal the role of the sign structure.

In order to examine the novel role of $\tau_c$, let us consider in the most simplified case of the 1D chain, i.e., with an open boundary condition. Here only a self-retracing path will contribute to the partition function in $Z_{t\text{-}J}$ (in order for the whole spin-hole configurations to return to the same ones in carrying out the trace in $Z_{t\text{-}J}$) such that $N_h^ \downarrow  [c]$ in $\tau_c$ is generally an even number for any closed path in Eq.~(\ref{ztj}). Namely $Z_{t\text{-}J}=  Z_{\sigma\cdot t\text{-}J}$ since $\tau_c=1$. Correspondingly, the eigen energies are also the same for the two models.

However, for such a 1D chain under open boundary condition, $\tau_c$ remains nontrivial for an open path, say, in the single-particle propagator Eq.~(\ref{gtj}). In fact, without the interference effect due to $\tau_c=1$ for a closed path, one may introduce a unitary transformation to ``gauge away'' the phase-string signs in the $t$-$J$ model \cite{Weng1997},
\begin{align}
  \label{Eq:PS_Tans}
   \E^{\I\hat{\Theta}}  \equiv \E^{-\I\sum_{i}n_i^h \hat{\Omega}_i}~,
\end{align}
where $\hat{\Omega}_i\equiv \sum_{l}\theta_i(l) n_{l\downarrow}$ with the statistical angle $\theta_i(l)$ satisfying
\begin{align}
  \label{}
  \theta_i(l) = \mathrm{Im}\ln(i-l) =
  \begin{cases} \pm\pi, & \text{if  } i<l,\\
	0, & \text{if  } i>l,
  \end{cases}
\end{align}
such that
\begin{align}
  \label{eq:ps}
\E^{-\I\hat{\Omega}_i}= \E^{\mp\I\pi\sum_{l>i}n_{l\downarrow}}\equiv (-1)^{\sum_{l>i}n_{l\downarrow}}~.
\end{align}
Here, $n_i^h$ and  $n_{l\downarrow}$ are the hole number and down-spin number operator, respectively.

Then, it is straightforward to show that the sign-full $t$-$J$ and sign-free $\sigma$$\cdot$$t$-$J$ models can be related by such a unitary transformation
\begin{align}
  \label{Eq:Trans}
  H_{t\text{-}J}  = \E^{\I\hat{\Theta}}   H_{\sigma\cdot t\text{-}J} \E^{-\I\hat{\Theta}}  .
\end{align}
Consequently, any eigenstate $|\Phi\rangle_{\sigma\cdot t\text{-}J}$ of the $\sigma$$\cdot$$t$-$J$ model, which has no ``sign problem", can be used to construct the corresponding eigenstate $|\Psi\rangle_{t\text{-}J}$ of the $t$-$J$ model of the same energy by
\begin{align}
  \label{eq:wf12}
  |\Psi\rangle_{t\text{-}J} = \E^{\I\hat{\Theta}}   |\Phi\rangle_{\sigma\cdot t\text{-}J}.
\end{align}
Although we shall focus on the single hole case below, the above construction is rigorous for an arbitrary hole concentration of the 1D chain under an open boundary condition \cite{Weng1997}.

\subsection{More detailed sign structure in wave functions}
\label{sec:sign_model}

One has seen above that a doped hole in a Mott insulator will generally induce a phase (sign) shift, which is many-body (dependent on the background spins) and irreparable or unrenormalizable (as $\mathcal{Z}[c]$, $\mathcal{W}[c] \ge 0$). The sign structure $\tau_c$ in Eq.~(\ref{tauc}) is obviously different from a conventional Fermi sign structure. It reflects a peculiar quantum ``memory'' effect of the hole moving on the quantum spin background. Namely it reflects a long-range entanglement between the charge and spin degrees of freedom.

In order to further identify the sign structure in the wave function, we start from the undoped case. At half-filling, both the $t$-$J$ model and $\sigma$$\cdot$$t$-$J$ model reduce to the same Heisenberg Hamiltonian $H_J$. According to Marshall\cite{Marshall1955}, the ground-state wave function of the Heisenberg model for a bipartite lattice is real in the Ising basis and satisfies a Marshall sign rule. This sign rule requires that the flip of a pair of antiparallel spins at nearest-neighbor sites will induce a sign change in the wave function, i.e., $\uparrow\downarrow$ $\rightarrow$ ($-1$) $\downarrow\uparrow$. If one introduces the so-called Marshall basis with the built-in Marshall sign by
 \begin{align}
  \label{eq:Marshall_basis}
  |\{s\}\rangle=(-1)^{N^\downarrow_A} c_{1s_1}^\dagger c_{2s_1}^\dagger \cdots c_{Ls_L}^\dagger |0\rangle,
\end{align}
where $L$ is the total site number or chain length and $N^\downarrow_A$ denotes the total number of down spins belonging to the sublattice $A$,  it is straightforward to verify that the off-diagonal matrix elements of $H_J$ are non-positive. Then the half-filling ground state of the Heisenberg model, denoted by $|\phi_0\rangle$, has non-negative coefficients in the Marshall basis according to the Perron-Frobenius theorem, namely, \begin{align}
  \label{eq:hfgs}
 |\phi_0\rangle=\sum_{\{s\}} c\left(\{s\}\right)|\{s\}\rangle,
\end{align}
with $c\left(\{s\}\right)\ge 0$. Thus $|\phi_0\rangle$ is indeed sign-free in the Marshall basis $\{|\{s\}\rangle\}$.

The single hole doping can be then realized by removing a spin-$\sigma$ ($\sigma=\pm1$) from the half filling singlet state, leading to a total spin $S^z_{\mathrm{tot}}=-\sigma/2$ single-hole state. Starting from the Marshall basis $\{|\{s\}\rangle\}$, one may further construct a new sign-free basis for the single-hole ground state of the $\sigma$$\cdot$$t$-$J$ mode (see Appendix~\ref{Appen:sign}): $\{(-\sigma)^i c_{i\sigma} |\{s\}\rangle\}$, where the sign factor $(-\sigma)^i $ comes from the Marshall sign associated with site $i$. Namely the ground state of the $\sigma$$\cdot$$t$-$J$ Hamiltonian can be written as
\begin{align}
  \label{eq:stJ_GSsign}
  |\Phi\rangle_{\sigma\cdot t\text{-}J} =\sum_{i,\{s\}} a(i,\{s\})\,  (-\sigma)^{i} c_{i\sigma} |\{s\}\rangle
\end{align}
with $a(i,\{s\})\ge0$. This is easy to understand as there is no nontrivial sign structure in the $\sigma$$\cdot$$t$-$J$ model, similar to the Heisenberg model at half-filling.

Then, according to Eq.~(\ref{eq:wf12}), the ground state of the $t$-$J$ model can be precisely constructed in terms of Eq.~(\ref{eq:stJ_GSsign}) as follows (cf. Appendix~\ref{Appen:sign}):
\begin{align}
  \label{eq:tJ_GSsign}
  |\Psi\rangle_{t\text{-}J}  =\sum_{i,\{s\}} a(i,\{s\})\, \E^{-\I \hat{\Omega}_i} (-\sigma)^i c_{i\sigma} |\{s\}\rangle ~,
\end{align}
where the many-body phase shift (phase string) operator $\E^{-\I \hat{\Omega}_i}$ is defined by Eq.~(\ref{eq:ps}).

\begin{figure}[tbp]
\begin{center}
\includegraphics[width=0.45\textwidth]{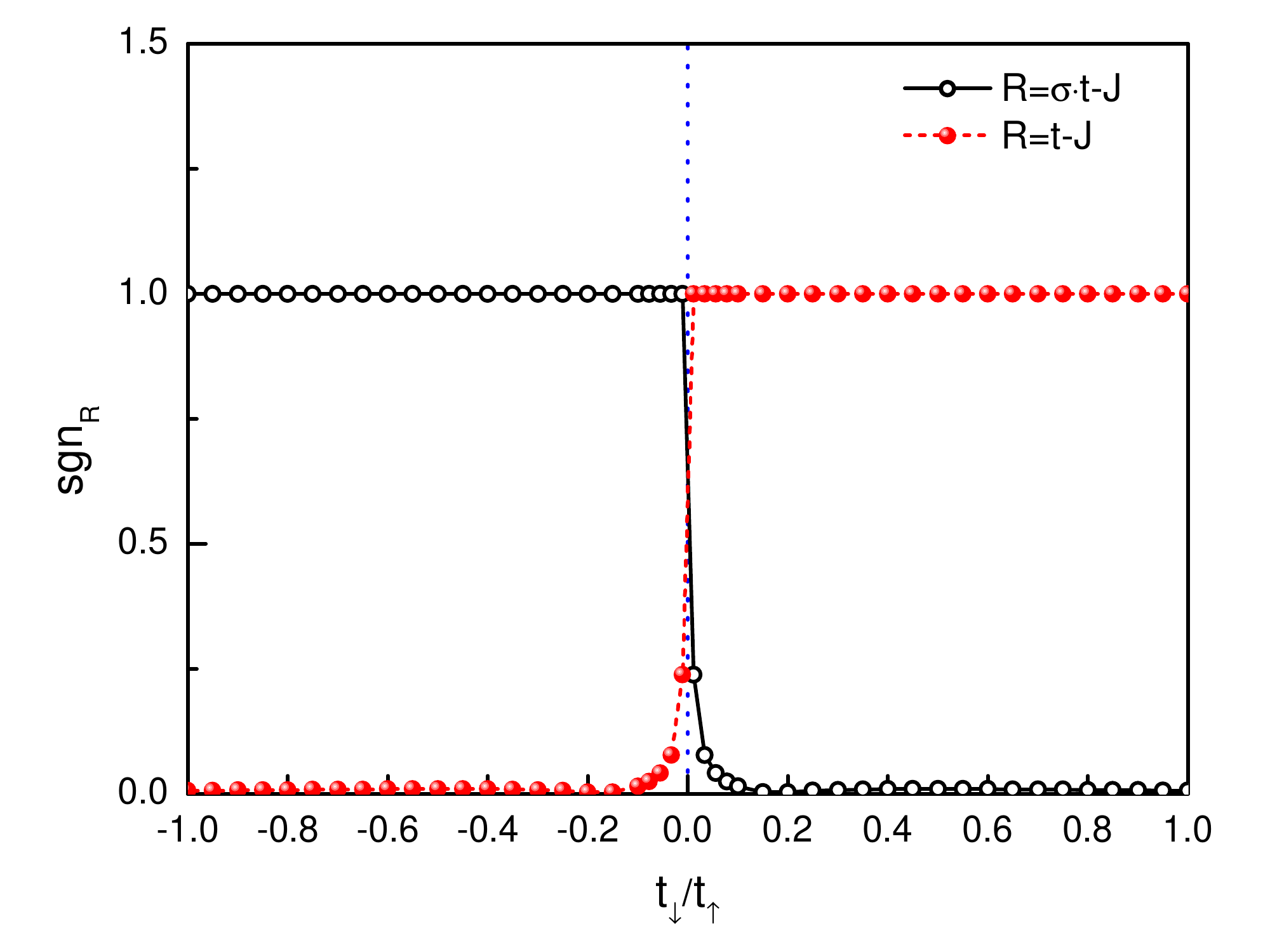}
\end{center}
\par
\renewcommand{\figurename}{Fig.}
\caption{(Color online) The intrinsic sign structure may be measured by a sign average $\operatorname{sgn}_R$ [defined in Eqs.~(\ref{eq:sign_ave}) and (\ref{eq:sign_ave_basis})] for the ground state of the $t_\uparrow$-$t_\downarrow$-$J$ model, which continuously interpolates the $t$-$J$ and $\sigma$$\cdot$$t$-$J$ models as a function of the ratio $t_\downarrow/t_\uparrow$ (see text). The results here are obtained by the exact diagonalization with a chain length $L=12$.  Here, $R$=$t$-$J$ and $R$=$\sigma$$\cdot$$t$-$J$ represent two sets of basis [see Eq. (\ref{eq:sign_ave_basis})] used to measure $\operatorname{sgn}_R$ of the ground state of $t_\uparrow$-$t_\downarrow$-$J$ model, respectively.  Clearly a critical point is indicated at $t_\downarrow/t_\uparrow=0$ where the sign structure shows a qualitative change in the ground state. A first-order-type of transition at this critical point is also to be confirmed by DMRG later.}
\label{Fig:SignAverage}
\end{figure}

\subsection{Breakdown of adiabatic continuity}
\label{sec:tup_tdown_J}

Two ground states, $ |\Psi\rangle_{t\text{-}J} $ and $ |\Phi\rangle_{\sigma\cdot t\text{-}J}$, differ by the phase string sign structure $\E^{\I\hat{\Theta}}$ according to Eq.~(\ref{eq:wf12}). In the following, we explicitly show the breakdown of the adiabatic continuity between them by examining the $t_\uparrow$-$t_\downarrow$-$J$ model introduced in Eq.~(\ref{Eq:tutdJmodel}), which interpolates the $t$-$J$ and $\sigma$$\cdot$$t$-$J$ models under an open boundary condition.

We define a sign average, motivated by Ref.~\onlinecite{Poilblanc2008}, to directly ``measure'' the sign structure of a generic wave function $|\psi\rangle$ in a given basis $\{|n_R\rangle\}$:
\begin{align}
  \label{eq:sign_ave}
  \operatorname{sgn}_R = \frac{\sum_{n_R} \operatorname{sgn}\left(\langle n_R|\psi\rangle\right) |\langle n_R|\psi\rangle|^2 }{\sum_{n_R} |\langle n_R|\psi\rangle|^2}.
\end{align}
The crucial feature in the above definition is that the signs are averaged according to the probability $|\langle n_R|\psi\rangle|^2$. If the coefficients $\langle n_R|\psi\rangle$ of the wave function in this particular basis have the same sign, then we have $\operatorname{sgn}_R=1$. Otherwise the coefficients with different signs cancel each other at least partially, resulting in a relatively small sign average.

For the models we focus on in this paper, the basis states $|n_R\rangle$ are chosen to be
\begin{align}
  \label{eq:sign_ave_basis}
  |n_R\rangle =
  \begin{cases}
  (-\sigma)^{i} c_{i\sigma} |\{s\}\rangle, & R=\sigma\mathrm{\cdot}t\text{-}J, \\
  \E^{-\I\hat{\Omega}_i} (-\sigma)^{i} c_{i\sigma} |\{s\}\rangle, & R=t\text{-}J,
  \end{cases}
\end{align}
which are the sign-free bases for the $\sigma$$\cdot$$t$-$J$ model and the $t$-$J$ model, respectively, as discussed above.

Based on the sign structure analysis there, one can show that the sign average $\operatorname{sgn}_{\sigma\cdot t\text{-}J}$ defined above for the $t_\uparrow$-$t_\downarrow$-$J$ model ($t_\uparrow=t$ is fixed to be positive) is
\begin{align}
  \label{eq:signaverage1}
  \operatorname{sgn}_{\sigma \mathrm{\cdot} t \text{-} J} =
  \begin{cases}
    1, & t_\downarrow/t_\uparrow<0, \\
    \langle \psi(-t_\downarrow/t_\uparrow) | \psi(t_\downarrow/t_\uparrow) \rangle, & t_\downarrow/t_\uparrow>0,
  \end{cases}
\end{align}
where $|\psi(t_\downarrow/t_\uparrow)\rangle$ is the ground state of the $t_\uparrow$-$t_\downarrow$-$J$ model with parameter $t_\downarrow/t_\uparrow$. A similar but reversed result is expected for the sign average $\operatorname{sgn}_{t\text{-}J}$.
The sign average $\operatorname{sgn}_{\sigma\mathrm{\cdot}t\text{-}J}$ equals to one exactly for $t_\downarrow/t_\uparrow<0$, because the basis $|n_{\sigma\mathrm{\cdot}t\text{-}J}\rangle$ is the sign-free basis in this parameter region. On the other hand, for $t_\downarrow/t_\uparrow>0$, the sign average measures the overlap between the two ground states for parameters $\pm t_\downarrow/t_\uparrow$. Therefore, a diminished sign average for $t_\downarrow/t_\uparrow = 0^+$ (see numerical results below) indicates the orthogonality of the ground states with parameters $t_\downarrow/t_\uparrow = 0^\pm$, i.e., the discontinuity in connecting the $\sigma$$\cdot$$t$-$J$ model and the $t$-$J$ model.

We perform an exact diagonalization of the $t_\uparrow$-$t_\downarrow$-$J$ model and calculate the sign average defined in Eqs.~(\ref{eq:sign_ave}) and (\ref{eq:sign_ave_basis}). The results on a chain with length $L=12$ are shown in Fig.~\ref{Fig:SignAverage}. More abrupt changes of the sign averages are expected for larger system sizes at the transition point, which is clearly at $t_\downarrow/t_\uparrow=0$.

\subsection{Distinct momentum structures}
\label{sec:sign_Zknk}

For the 1D chain under an open boundary condition, the single-hole ground state energy does not depend on the sign structure, due to the absence of the nontrivial phase string $\tau_c$ for closed paths as to be verified by a DMRG calculation later. Nevertheless,  as shown above, the two ground states cannot be smoothly connected, which means that the phase string, as captured by $\E^{-\I\hat{\Omega}_i} $ in Eq.~(\ref{eq:tJ_GSsign}), still leads to completely different physical properties of the ground states\cite{Weng1997,Suzuura1997,Nagaosa1998}, even in the absence of the interference effect.

In the following, we specifically examine the characteristic momenta in terms of the quasiparticle spectral weight $Z_k$ and the momentum distribution $n_{k\alpha}$ for the two models. Other properties related to the sign structure, such as the total spin of the ground state and the ordering of energy levels for the $t$-$J$ chain, can be found in Ref.~\onlinecite{Wang2014}.

Firstly we start with the ground state wave function Eq.~(\ref{eq:stJ_GSsign}) of the $\sigma$$\cdot$$t$-$J$ chain. The quasiparticle spectral weight is defined as $Z_k=|a_k|^2$, where $a_k$ is the overlap between the ground state $|\Phi\rangle_{\sigma\cdot t\text{-}J}$ and the Bloch-like state $|k\rangle = \frac{1}{\sqrt L}\sum_i \E^{-\I k i} c_{i\sigma} |\phi_0\rangle$ constructed from the ground state of the Heisenberg model $|\phi_0\rangle$ by removing an electron:
\begin{align}
  \label{eq:a_k}\nonumber
  a_k &\equiv \langle k | \Phi \rangle_{\sigma\cdot t\text{-}J} \\
  &= \frac{1}{\sqrt L} \sum_{i,\{s\}} a(i,\{s\}) \E^{\I k i} (-\sigma)^{i} \langle \phi_0| c_{i\sigma}^\dagger c_{i\sigma} |\{s\}\rangle.
\end{align}
Because of the fact that $a(i,\{s\}) \ge 0$ and $\langle \phi_0| c_{i\sigma}^\dagger c_{i\sigma} |\{s\}\rangle \ge 0$, it is straightforward to see that the quasiparticle weight $Z_k$ must be peaked at $k=0$ ($k=\pi$) if $\sigma=\downarrow$ ($\sigma=\uparrow$). Note that here $\sigma$ denotes the spin removed from the half-filling in the one-hole state, i.e., the ground state $| \Phi \rangle_{\sigma\cdot t\text{-}J}$ has total spin $S^z_\mathrm{tot} = -\sigma/2$.

Similarly, the momentum distribution $n_{k\alpha}$ ($\alpha=\sigma,\bar\sigma$) of the $\sigma$$\cdot$$t$-$J$ ground state wave function is given by
\begin{align}
  \label{}\nonumber
  n_{k\alpha} &= \langle\Phi| c_{k\alpha}^\dagger c_{k\alpha} |\Phi\rangle_{\sigma\cdot t\text{-}J}\\
  &= \frac{1}{2} - \frac{1}{L}\delta_{\alpha\sigma} + \frac{1}{L} \sum_{i\neq j} \E^{-\I k(j-i)} \langle\Phi| c_{i\alpha}^\dagger c_{j\alpha} |\Phi\rangle_{\sigma\cdot t\text{-}J},
\end{align}
where
\begin{align}
  \label{eq:n_k}\nonumber
  \langle\Phi| c_{i\alpha}^\dagger c_{j\alpha} |\Phi\rangle_{\sigma\cdot t\text{-}J}
  &\equiv \sum_{\{s\},\{s'\}} a(j,\{s'\}) a(i,\{s\}) (-\sigma)^{i-j} (-1)\\
  &\quad\ \times
  \begin{cases}
    \langle\{s'\}| n_{i\sigma} n_{j\sigma} |\{s\}\rangle, & \alpha=\sigma,\\
    \langle\{s'\}| S_i^{\mp} S_j^{\pm} |\{s\}\rangle, & \alpha=\bar\sigma.
  \end{cases}
\end{align}
For $\alpha=\sigma$ (again $\sigma$ denotes the spin removed in the single hole case), the momentum distribution $n_{k\sigma}$ will show a sharp dip (relative to the half-filling) at $k=0$ ($k=\pi$) if $\sigma=\downarrow$ ($\sigma=\uparrow$) because of $a(i,\{s\}) \ge 0$. On the other hand, if $\alpha=\bar\sigma$, there is an another Marshall sign contribution $(-1)^{i+j}$ from $\langle\{s'\}| S_i^{\mp} S_j^{\pm} |\{s\}\rangle$. Hence the momentum distribution $n_{k\bar\sigma}$ should exhibit a sharp dip at $k=\pi$ ($k=0$) at $\sigma=\downarrow$ ($\sigma=\uparrow$).  These characteristic momenta manifested in $Z_k$ and $n_k$ for the $\sigma$$\cdot$$t$-$J$ chain are summarized in the Table~\ref{tab:peaks}.

\begin{table}[h]
\centering
\caption{Characteristic momenta determined by $Z_k$ and $n_{k\alpha}$, which are dependent on the detailed sign structure in the ground states of $\sigma$$\cdot$$t$-$J$ and $t$-$J$ models, respectively. Here the single hole is doped into the half-filling ground state by removing a spin of $\sigma$.}

\label{tab:peaks}
\begin{tabular}{|c|c|c|c|c|}
\hline
models                                    & $\sigma$     & $Z_k$   & $n_{k\uparrow}$          & $n_{k\downarrow}$      \\ \hline
\multirow{2}{*}{$\sigma\mathrm{\cdot}t$-$J$} & $\uparrow$   & $k=\pi$ & \multirow{2}{*}{$k=\pi$} & \multirow{2}{*}{$k=0$} \\ \cline{2-3}
                                          & $\downarrow$ & $k=0$   &                          &                        \\ \hline
\multirow{2}{*}{$t$-$J$}                  & $\uparrow$   & \multicolumn{3}{c|}{\multirow{2}{*}{$k=\pm\pi/2$}}          \\ \cline{2-2}
                                          & $\downarrow$ & \multicolumn{3}{c|}{}                                       \\ \hline
\end{tabular}
\end{table}

Then we similarly examine the case for the $t$-$J$ model. The quasiparticle weight $Z_k=|a_k|^2$ can be calculated similarly to Eq.~(\ref{eq:a_k}), with $\langle \phi_0| c_{i\sigma}^\dagger c_{i\sigma} |\{s\}\rangle$ replaced by $\langle \phi_0| \E^{-\I \hat{\Omega}_i} c_{i\sigma}^\dagger c_{i\sigma} |\{s\}\rangle$.
Since the ground state $|\phi_0\rangle$ of the Heisenberg spin chain has an antiferromagnetic correlation, the phase string effect of $\E^{-\I \hat{\Omega}_i}$ will cause an ``unrenormalizable phase shift'' \cite{Anderson90PRL,Weng1991,Weng1992,Ren-Anderson1993}, whose leading order contribution is given by $ \E^{\pm\I(\pi/2)i}$, transforming the characteristic momentum at $k=0$ or $\pi$ of the $\sigma$$\cdot$$t$-$J$ model to $k=\pm\pi/2$ in the $t$-$J$ model. The fluctuation effect of $\E^{-\I \hat{\Omega}_i}$ around $ \E^{\pm\I(\pi/2)i}$ will further contribute to the power of the vanishing quasiparticle weight at large $L$ limit as to be discussed in the DMRG and VMC calculations later.

\begin{figure}[tbp]
\begin{center}
\includegraphics[width=0.4\textwidth]{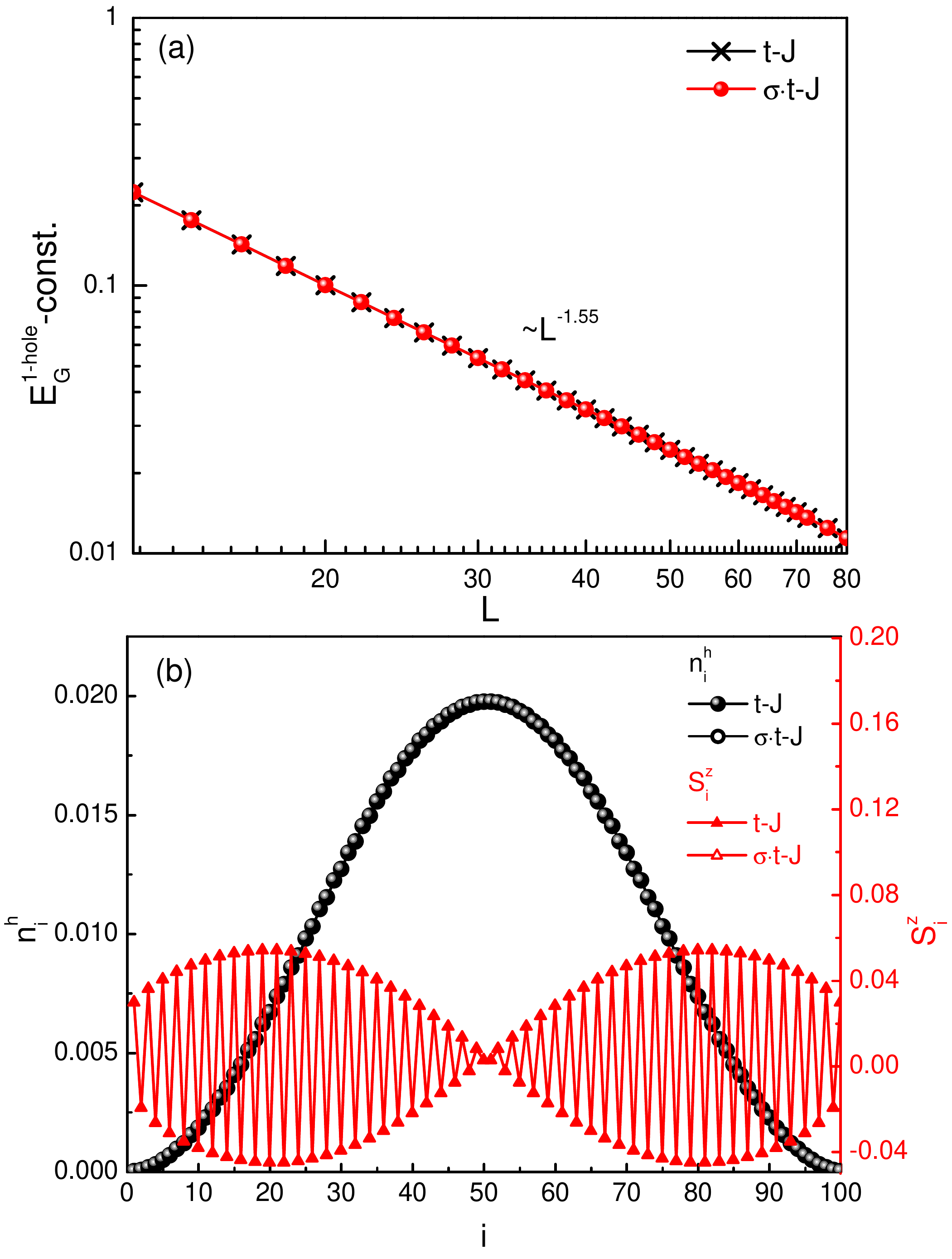}
\end{center}
\par
\renewcommand{\figurename}{Fig.}
\caption{(Color online) (a) The same one-hole ground-state energy $E_{G}^{\text{1-hole}}$ vs, the chain length $L$ for the $t$-$J$ and $\sigma$$\cdot$$t$-$J$ models due to the absence of the interference of the phase string under an open boundary condition; (b) The charge and spin density distributions are also the same for both models. Note that the spin-charge separation is clearly indicated in such a finite-size system. }
\label{Fig:OBC_E0}
\end{figure}

\begin{figure}[tbp]
\begin{center}
\includegraphics[height=3.8in,width=2.8in]{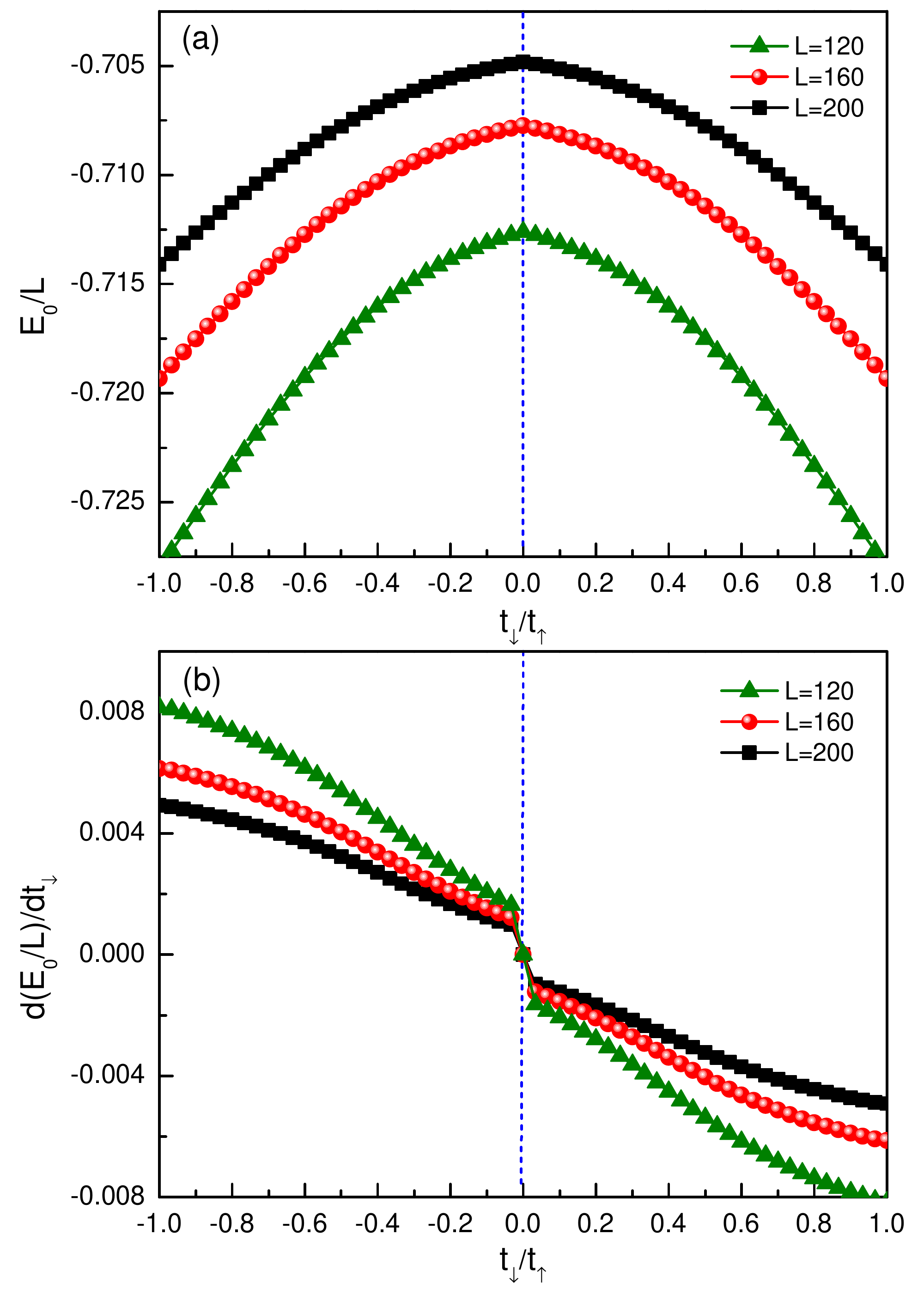}
\end{center}
\par
\renewcommand{\figurename}{Fig.}
\caption{(Color online) The ground state energy $E_0$ of the $t_\uparrow$-$t_\downarrow$-$J$ model as a function of $t_\downarrow/t_\uparrow$ [(a)], which exhibits a singularity at $t_\downarrow/t_\uparrow=0$ as indicated by the first order derivative of ground state energy [(b)]. It implies the breakdown of adiabatic continuity between the ground states of the $t$-$J$ and $\sigma$$\cdot$$t$-$J$ models at $t_\downarrow/t_\uparrow=1$ and $t_\downarrow/t_\uparrow=-1$, respectively, and is consistent with the sign structure analysis shown in Fig.~\ref{Fig:SignAverage}.}
\label{Fig:TupTdownModel}
\end{figure}
\begin{figure}[tbp]
\begin{center}
\includegraphics[height=5.6in,width=2.8in]{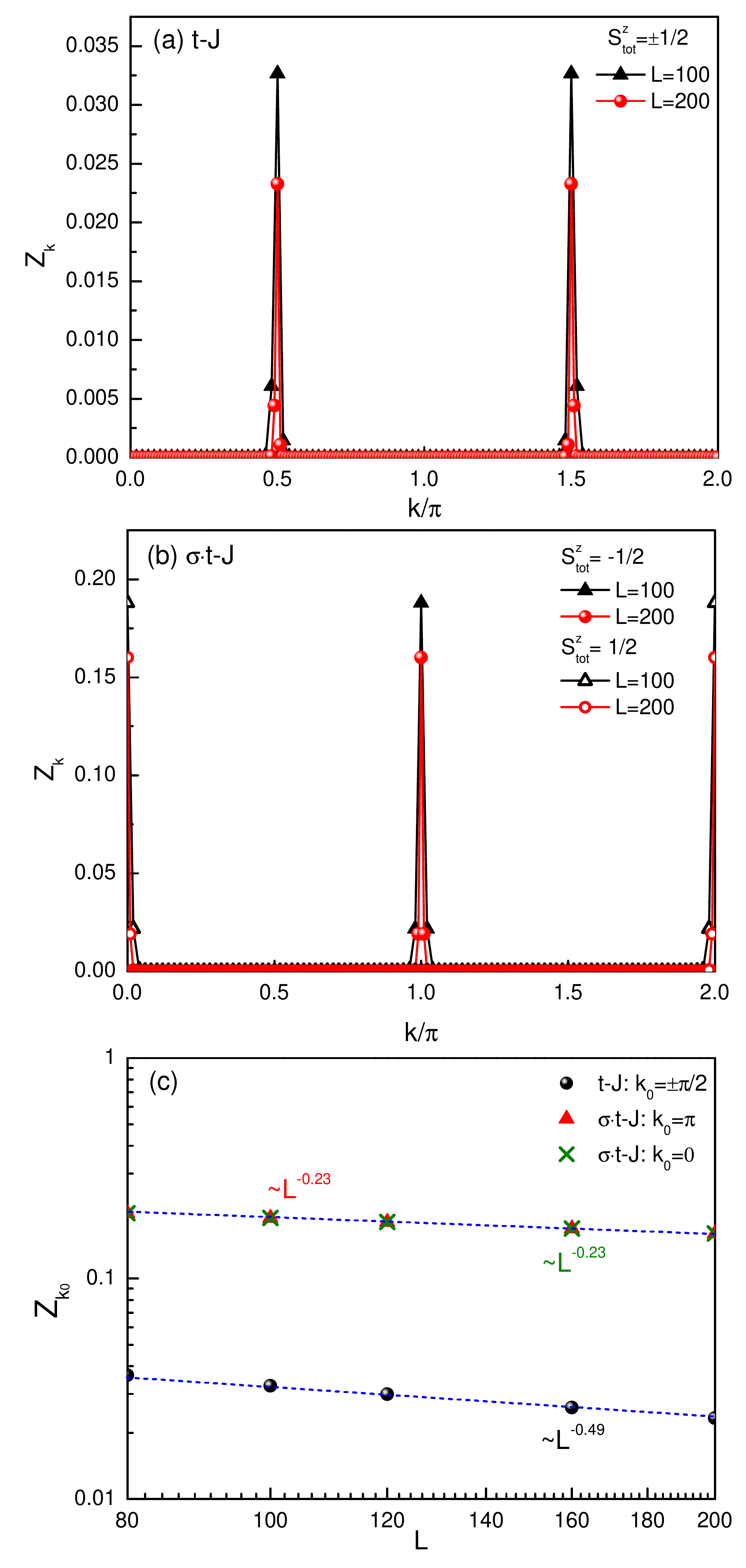}
\end{center}
\par
\renewcommand{\figurename}{Fig.}
\caption{(Color online)  The distinct characteristic momenta $k_0$'s as determined by the peaks of the single-particle spectrum weight $Z_k$. (a) The one-hole-doped $t$-$J$ chain; (b) The one-hole-doped $\sigma$$\cdot$$t$-$J$ chain. Here the single hole state is realized by removing an up-spin or down-spin, with $S^z_{tot}=-1/2$ and $S^z_{tot}=1/2$, respectively; (c) $Z_{k_0}$ vanishes in the thermodynamic limit in a power law fashion $L^{-\alpha}$,  with $\alpha \simeq 0.49$ for the $t$-$J$ and $\alpha \simeq 0.23$ for the $\sigma$$\cdot$$t$-$J$ models, respectively. }
\label{Fig:Zk}
\end{figure}
\begin{figure}[tbp]
\begin{center}
\includegraphics[height=5.6in,width=2.8in]{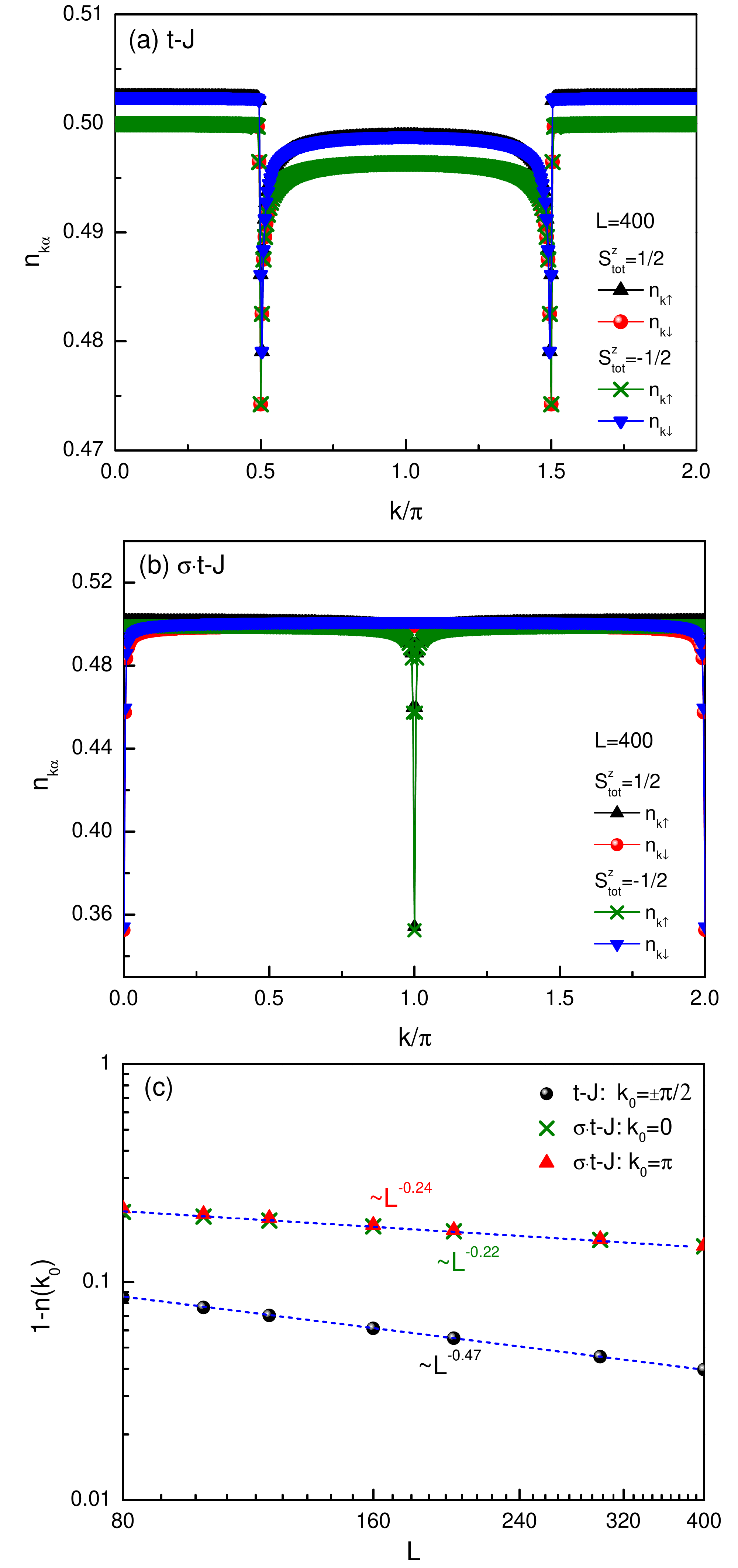}
\end{center}
\par
\renewcommand{\figurename}{Fig.}
\caption{(Color online) The momentum distributions, $n_{k\uparrow}$ and $n_{k\downarrow}$, for the $t$-$J$ (a) and the $\sigma$$\cdot$$t$-$J$ (b) models. $S^z_{tot}=\pm 1/2$ denotes the total spin of the one-hole state. (c) The peak of the hole momentum distribution $1-n(k_0)\equiv 1- n_{k_0\uparrow} - n_{k_0\downarrow}$ scales with $L^{-\alpha}$, with $\alpha \simeq 0.47$ for the $t$-$J$ model while $\alpha \simeq 0.22$ ($\alpha \simeq 0.24$) for the peaks located at $k_0=0$ ($k_0=\pi$) in the $\sigma$$\cdot$$t$-$J$ model.}
\label{Fig:1-nk}
\end{figure}

For the momentum distribution, one finds an additional phase string contribution $\E^{-\I \hat{\Omega}_i}\E^{\I \hat{\Omega}_j}$ in Eq.~(\ref{eq:n_k}), which in the leading order approximation, is given by $\E^{\pm\I (\pi/2)(j-i)}$. The phase string effect again shifts the momentum by $\pi/2$, leading to sharp dips of $n_{k\alpha}$ at $k=\pm\pi/2$. All of the above theoretical predictions of the peak/dip positions of $Z_k$ and $n_{k\alpha}$, based on the sign structures of the $\sigma$$\cdot$$t$-$J$ model and the $t$-$J$ model, are given in Table~\ref{tab:peaks}, which will be compared to the DMRG and VMC results in the following sections.

\section{DMRG results}
\label{sec:DMRG}

The present 1D problem can be accurately studied by the DMRG simulation\cite{DMRG1992}. Some numerical details are as follows. We shall make use of the conserved quantum numbers, i.e., the total particle number $N=L-1$ and total spin $S^z$ , and fix $t/J=3$ throughout the paper, setting $J$ as the unit of energy. The simulation ensures the truncation error of the order or less than $10^{-10}$  by keeping sufficient number of states (200$\sim$6000) and performing at least 40 sweeps for different boundary conditions. By calculating the quasiparticle weight, we obtain the undoped ground state first,  and then perform enough sweeps to get the one hole ground state by removing an electron in the center. The off-diagonal measurement ${Z_j} = \langle \Psi _\text{0-hole} |c_{j\sigma }^\dag | \Psi _\text{1-hole}\rangle $ are made after obtaining the converged wavefunction.

\subsection{Phase string effect under open boundary condition}
\label{Sec:OBC}

Since the single hole doped $t$-$J$ and $\sigma$$\cdot$$t$-$J$ chains can be connected by a unitary transformation [See. Eq.~(\ref{Eq:Trans})] under an open boundary condition, the
eigen energies are the same  for these two models. Indeed, the ground state energies calculated by the DMRG simulation are shown to coincide precisely in Fig.~\ref{Fig:OBC_E0} (a), where the one-hole energy $E_{G}^{\text{1-hole}}$ is defined by
\begin{equation}
E_{G}^{\text{1-hole}}=E_{0}^{\text{1-hole}}-E_{0}^{\text{0-hole}},
 \end{equation}
where $E_{0}^{\text{1-hole}}$ and $E_{0}^{\text{0-hole}}$ represent the ground-state energies of the one-hole-doped and half-filled cases, respectively.
Similarly, the physical quantities involving the local diagonal operators, like the hole density and spin distribution, $n^h_i  =1-n_i$ and $S^z_i$, are also the same on the ground states of $t$-$J$ and $\sigma$$\cdot$$t$-$J$ chains as shown in Fig.~\ref{Fig:OBC_E0} (b).

However, in Sec.~\ref{sec:tup_tdown_J}, we have shown that these two ground states, $ |\Psi\rangle_{t\text{-}J} $ and $ |\Phi\rangle_{\sigma\cdot t\text{-}J}$, differing by the phase string sign structure $\E^{-\I \hat{\Omega}_i}$, cannot be smoothly connected by the $t_\uparrow$-$t_\downarrow$-$J$ model Eq.~(\ref{Eq:tutdJmodel}), and there is an abrupt change of the sign average (cf. Fig.~\ref{Fig:SignAverage}) at $t_\downarrow/t_\uparrow=0$.

In Fig.~\ref{Fig:TupTdownModel} (a), the ground state energy $E_0$ of the $t_\uparrow$-$t_\downarrow$-$J$ model computed by DMRG is presented as a function of the ratio $t_\downarrow/t_\uparrow$. $E_0$ shows a singularity at $t_\downarrow/t_\uparrow=0$ by a sharp jump in the first-order derivative curve [see Fig.~\ref{Fig:TupTdownModel} (b)]. It indeed implies the breakdown of the adiabatic continuity between the two ground states by a first-order-type of transition, similar to Fig.~\ref{Fig:SignAverage} based on the sign structure.

Even though there is no closed path for the interference effect of phase string to take place here, the nontrivial effect of the sign structure will still play a critical role in determining the quasiparticle weight $Z_{k}$ and the momentum distribution $n_{k\uparrow}$ and $n_{k \downarrow}$ as predicted in Sec.~\ref{sec:sign_Zknk}.

Figure~\ref{Fig:Zk} shows the quasiparticle weight $Z_k$ obtained by DMRG. Different peak position $k_0$'s of $Z_k$ here indicate that the momentum structures are totally different in the two models: i.e., $k_0=\pm \pi/2$ in the $t$-$J$ model [Fig.~\ref{Fig:Zk} (a)] vs. $k_0=\pi$ and $k_0=0$ for $S^z_{tot}=-1/2$ and $S^z_{tot}=1/2$, respectively, in the $\sigma$$\cdot$$t$-$J$ model [Fig.~\ref{Fig:Zk} (b)]. These characteristic momentum peak positions in $Z_{\mathbf{k}}$ are consistent with the exact analysis in Sec.~\ref{sec:sign_Zknk} (cf. Table~\ref{tab:peaks}), where different sign structures are shown to be responsible for the distinction.

Furthermore, a power-law decay of $Z_k$ at $k_0$'s in a fashion of $Z_k\sim L^{-0.49}$ is found for the $t$-$J$ model [cf. Fig.~\ref{Fig:Zk} (c)]. By contrast,  $Z_{k_0}\sim L^{-0.23} $ is identified for the $\sigma$$\cdot$$t$-$J$ model. The disappearance of $Z_k$ at $L\rightarrow \infty$ is usually called the Luttinger liquid behavior.

For the $\sigma$$\cdot$$t$-$J$ model, the vanishing $Z_{k_0}$ may be purely attributed to the fact that the spin-1/2 associated with the doped hole can move away to infinite along the chain, i.e., the so-called spin-charge separation as the spin is gapless in the 1D chain. The charge and spin density distributions in Fig.~\ref{Fig:OBC_E0} (b) have clearly illustrated this. In fact, if one artificially turns on a gap in the spin chain, then a spin-charge recombination may re-take place in Fig.~\ref{Fig:OBC_E0} (b) and consequently a sharp peak with a finite $Z_{k_0}$, which corresponds to a conventional Bloch state, could be recovered.

For the $t$-$J$ model, the decay of $Z_{k_0}$ as a function of $L$ is steeper, because the phase string $\E^{-\I \hat{\Omega}_i}$ will fluctuate around $\E^{\pm\I(\pi/2)i}$ to contribute to an additional power-law decay (cf. Sec.~\ref{sec:sign_Zknk}). Here the phase string sign structure plays the role of many-body phase shift. It determines not only the total momentum of the one-hole ground state, but also the non-Fermi-liquid behavior through its fluctuation.

Similarly the distinct momentum structure is also manifested in the momentum distribution $n_{k\alpha}$ of the electrons as predicted in Table~\ref{tab:peaks}. As presented in Figs.~\ref{Fig:1-nk} (a) and (b), one sees that two sharp peaks (dips) of  $n_\uparrow$ or $n_\downarrow$ at $k_0=\pm \pi/2$ for the $t$-$J$ model, whereas at $k_0=\pi$ and $k_0=0$ for the $\sigma$$\cdot$$t$-$J$ model. Moreover, the height of the peak of the momentum distribution $1-n(k_0)$ [$n(k_0)\equiv n_{k_0\uparrow} + n_{k_0\downarrow}$] decays in a power-law fashion as shown in Fig.~\ref{Fig:1-nk} (c). It is again consistent with that the vanishing quasiparticle weight in the thermodynamic limit is much quicker for the $t$-$J$ model because of the phase string factor.

\subsection{Phase string interference under periodic boundary condition}

The phase string sign structure in the $t$-$J$ model can be ``gauged away'' by performing the phase-string transformation Eq.~(\ref{Eq:PS_Tans}) under an open boundary condition. Correspondingly the ground-state energies $E_{G}^{\text{1-hole}}$ are the same for the models with and without phase string signs, as shown in Fig.~\ref{Fig:OBC_E0}.

\begin{figure}[tbp]
\begin{center}
\includegraphics[height=5.6in,width=2.8in]{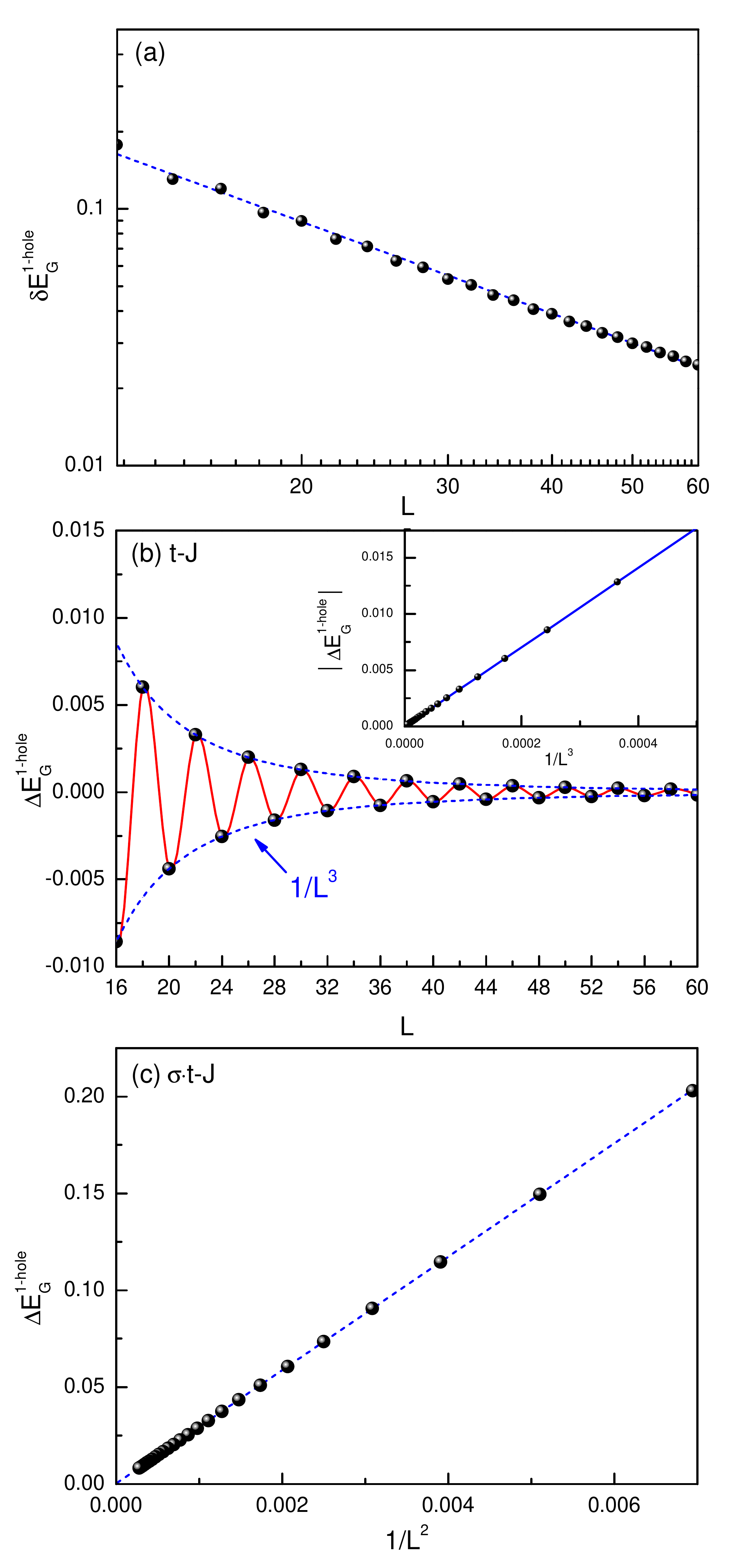}
\end{center}
\par
\renewcommand{\figurename}{Fig.}
\caption{(Color online) (a) The ground state energies are no longer the same for the $t$-$J$ and $\sigma$$\cdot$$t$-$J$ models under a periodic boundary condition. Their energy difference $\delta E_{G}^{\text{1-hole}}$ [cf. Eq.~(\ref{eq:deltaE})] is shown to vanish $\sim L^{-1}$ at large $L$;  The ground state energy change under inserting a flux into the 1D ring [cf. Eq.~(\ref{dE})], $\Delta E_{G}^{\text{1-hole}}$, for the $t$-$J$ model (b) and the $\sigma$$\cdot$$t$-$J$ model (c). $\Delta E_{G}^{\text{1-hole}}$ oscillates and decays in a power-law fashion ($\sim L^{-3}$) for the $t$-$J$ chain while it is non-oscillating and proportional to $1/L^2$ for the $\sigma$$\cdot$$t$-$J$ chain (Ref.\onlinecite{ZZ2013}).}
\label{Fig:PBC_E0}
\end{figure}

However, if a periodic boundary condition is imposed, such a unitary transformation no longer exists. In this case, one expects the quantum interference effect to take place as the hole can circumvent the closed 1D ring of a finite size. Consequently the energy degeneracy of the $t$-$J$ and $\sigma$$\cdot$$t$-$J$ models gets lifted due to the phase string sign structure:
\begin{align}
  \label{eq:deltaE}\nonumber
  \delta E_{G}^{\text{1-hole}} &\equiv E_{G}^{\text{1-hole}} (t\text{-}J) - E_{G}^{\text{1-hole}} (\sigma\mathrm{\cdot}t\text{-}J)\\\nonumber
  &= \lim_{\beta\to\infty} -\frac{1}{\beta} \ln \left( \frac{Z_{t\text{-}J}} {Z_{\sigma\mathrm{\cdot}t\text{-}J}} \right)\\\nonumber
  &= \lim_{\beta\to\infty} -\frac{1}{\beta} \ln \left( \frac{ \sum_{c} \tau_c \mathcal{Z}[c] } { \sum_c \mathcal{Z}[c] } \right)\\
  &\equiv \lim_{\beta\to\infty} -\frac{1}{\beta} \ln \langle \tau_c \rangle_{\mathcal{Z}}~,
\end{align}
which is non-vanishing as long as the phase interference takes place such that $\langle \tau_c \rangle_{\mathcal{Z}}<1$ (instead of $\langle \tau_c \rangle_{\mathcal{Z}}=1$ for the open boundary condition).
The detailed discussion of the sign structures of the two models under periodic boundary condition can be found in Appendix~\ref{Appen:sign_PBC}.
The energy difference is shown in Fig.~\ref{Fig:PBC_E0} (a), where the scaling law is also altered as compared to Fig.~\ref{Fig:OBC_E0} (a).

\begin{figure}[tbp]
\begin{center}
\includegraphics[width=8.3cm]{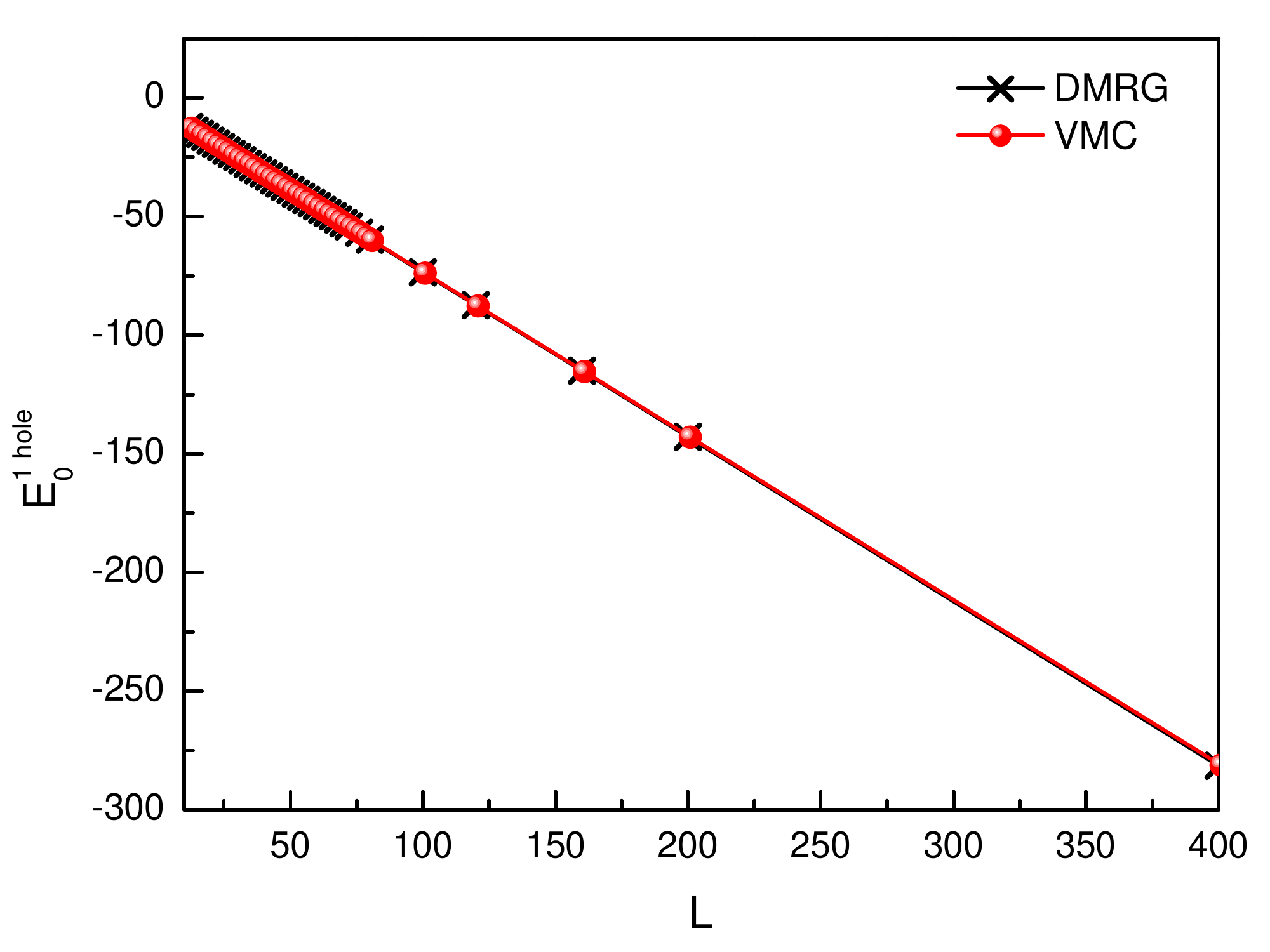}
\end{center}
\par
\renewcommand{\figurename}{Fig.}
\caption{(Color online) The one-hole ground-state energy for the variational wave function Eq.~(\ref{eq:wf1}) (at $t/J=3$) calculated by VMC (red dots) is in excellent agreement with the DMRG (black crosses).}
\label{Fig:E_compare}
\end{figure}

\begin{figure}[tbp]
\begin{center}
\includegraphics[height=5.6in,width=2.8in]{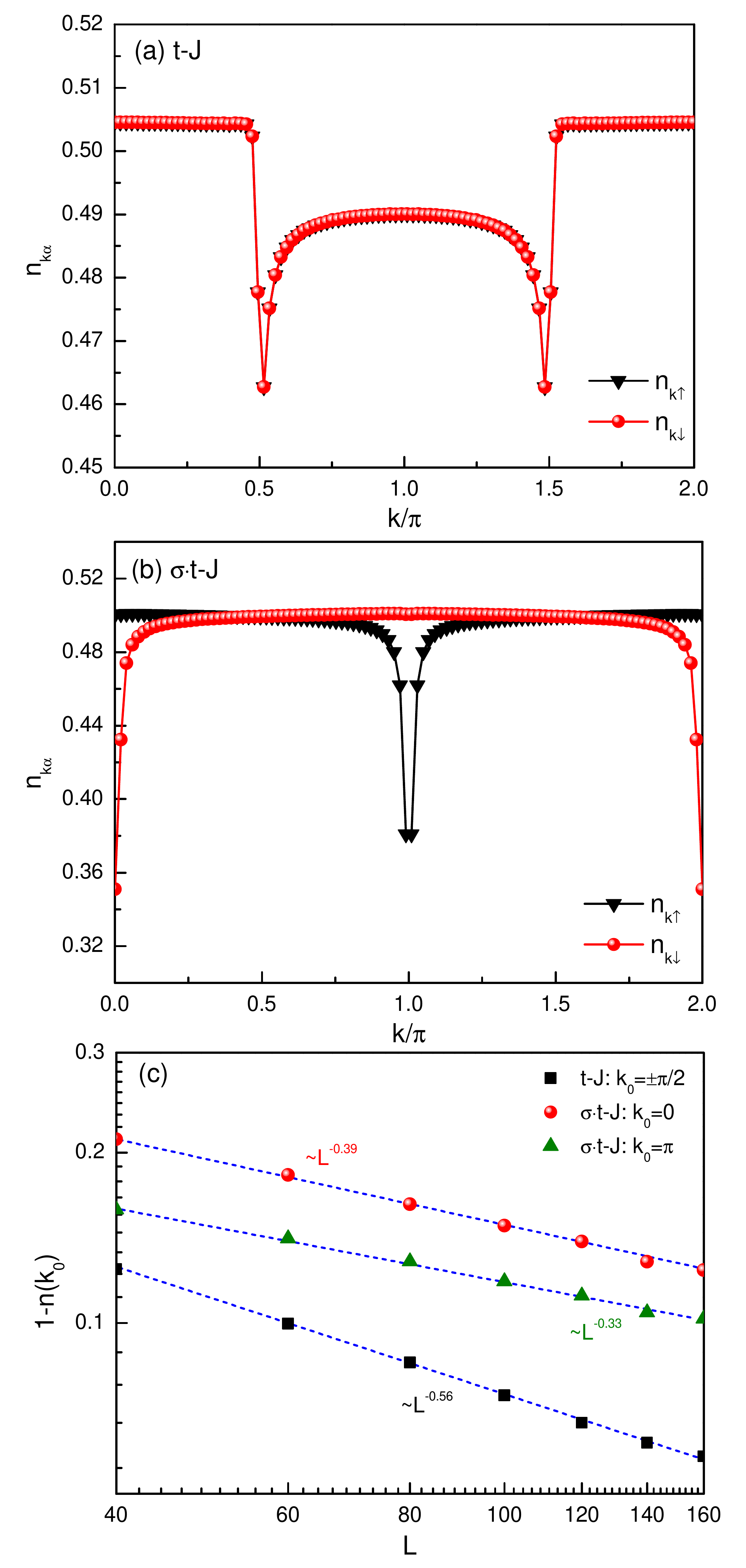}
\end{center}
\par
\renewcommand{\figurename}{Fig.}
\caption{(Color online) Momentum distribution $n_{k\alpha}$ calculated based on the variational wave function Eq.~(\ref{eq:wf1}) for $t$-$J$ model (a) and Eq.~(\ref{eq:wf2}) for the $\sigma$$\cdot$$t$-$J$ model (b), by VMC with length $L=101$ at $t/J=3$. The sharp dip positions are the same as predicted in Table~\ref{tab:peaks} as well as the DMRG results in Fig.~\ref{Fig:1-nk}; (c) The peak of the hole momentum distribution $1-n(k_0)=1- n_{k_0\uparrow} - n_{k_0\downarrow}$ vanishes in a power law fashion, $L^{-\alpha}$, in the thermodynamic limit.}
\label{Fig:MC}
\end{figure}

Since the phase string sign structure arises from the hole doping, one may detect it directly by measuring the behavior of the charge. For this purpose, one may insert a magnetic flux $\Phi$ threading through the 1D ring and then compute the ground state energy difference between $\Phi=\pi$ and $0$, i.e.,
\begin{equation}\label{dE}
 \Delta E_{G}^{\text{1-hole}}\equiv E_{G}^{\text{1-hole}}(\Phi=\pi )-E_{G}^{\text{1-hole}}(\Phi=0),
 \end{equation}
 which measures solely the charge sector that couples to the magnetic flux \cite{ZZ2013}.

Figure~\ref{Fig:PBC_E0} (b) shows the DMRG result of $\Delta E_{G}^{\text{1-hole}}$ for the single hole doped $t$-$J$ model, which oscillates and decays in a power-law fashion $L^{-3}$ [cf. the inset of Fig.~\ref{Fig:PBC_E0} (b)]. The period of the oscillation in $\Delta E_{G}^{\text{1-hole}}$ is 4 lattice constant, which matches with the momentum $k_0=\pm\pi/2$.
Such an oscillation itself reflects the phase string signs, which disappears once the sign structure is turned off in the $\sigma$$\cdot$$t$-$J$ model as shown in Fig.~\ref{Fig:PBC_E0} (c), where $\Delta E_{G}^{\text{1-hole}}$ is proportional to $1/L^2$  without any oscillation.

\section{Variational wave function}
\label{sec:wf}

So far we have established, via the exact analysis and the DMRG simulation in Sections~\ref{sec:model} and \ref{sec:DMRG}, that the phase string sign structure is essential in understanding the correct momentum structure and non-Bloch-wave behavior of a single hole injected into an AF Heisenberg spin chain.

In the following,  we shall further construct a variational ground state wave function based on the identified sign structure. By a VMC calculation, we show that such a trial wave function well reproduces the DMRG results, in which the role of the phase string is explicitly illustrated.

\subsection{Wave function incorporating the correct sign structure}

The general single-hole ground state of the $t$-$J$ chain under the open boundary condition may be formally written as
\begin{align}
  \label{eq:vwf0}
  |\Psi\rangle_{t{\text-}J} = \sum_i \varphi_h(i)\,\left[ \E^{-\I\hat\Omega_i}\hat{P}_{i} \right] c_{i\sigma}\, |\phi_0\rangle~,
\end{align}
where an electron of spin $\sigma$ is annihilated by $c_{i\sigma}$ from the singlet ground state $ |\phi_0\rangle$ at half-filling. Besides a single-hole wave function $\varphi_h(i)$ which is presumably a smooth function, the non-Bloch-wave part of the wave function involves the nontrivial sign structure $\E^{-\I\hat\Omega_i}$ identified in Sec.~\ref{sec:exactsign}, with $ \hat{P}_{i}$ denoting the rest of non-sign-related spin cloud (spin-polaron), in response to the bare hole state $c_{i\sigma}\, |\phi_0\rangle$.

According to Eq.~(\ref{eq:wf12}), the corresponding variational ground state for the $\sigma\cdot$ $t$-$J$ model is given by
\begin{align}
 \label{eq:vwf1}
 |\Phi\rangle_{\sigma\cdot t\text{-}J} = \sum_i \varphi_h(i)\hat{P}_{i}\, c_{i\sigma}\,  |\phi_0\rangle~,
\end{align}
where the phase string sign structure is gauged away, but $ \hat{P}_{i}$ is still present.

The wave function Eqs.~(\ref{eq:vwf0}) and (\ref{eq:vwf1}) are similar to that previously proposed for the single-hole ground states of a two-leg ladder case \cite{ladder2015}. The latter case is further simplified because the half-filling two-leg ladder is fully gapped such that the spin-polaron effect described by $ \hat{P}_{i}$ is not essential for the long-wavelength, low-energy physics, and thus is neglected in Ref.~\onlinecite{ladder2015}. Then $|\Phi\rangle_{\sigma\cdot t\text{-}J}$ there becomes a Bloch-wave state with $ \varphi_h(i)\propto \E^{\I k i}$.

However, in the present 1D chain case, the spin background is gapless with quasi-long-range spin correlations in $|\phi_0\rangle$. In the following, we shall see that $\hat{P}_{i}$ here will be related to the correction due to spin-charge separation in the $\sigma$$\cdot$$t$-$J$ model: the spin-1/2 associated with the doped hole can move away from the charge as a gapless spinon, which merely reflects the fact that the free gapless spinon already exists at half-filling. $\hat{P}_{i}$ will be approximately datermined based on the so-called squeezed spin chain construction.

\subsection{Squeezed spin chain construction}

The ground state of the doped Hubbard model at $U\gg t$ or the $t$-$J$ model at $J\ll t$ can be well described by the so-called squeezed spin chain approximation \cite{Ogata1990,Weng1991,Weng1992,Shiba1992,Weng1994,Kruis2004}, based on which the precise phase string sign structure has been first identified in Refs. \onlinecite{Weng1991,Weng1992,Weng1994,Weng1997}.

According to the squeezed spin chain approximation \cite{Ogata1990,Weng1991,Weng1992,Shiba1992,Weng1994,Kruis2004}, one may construct the single-hole-doped ground state as follows.  Starting from the half-filling ground state $|\phi_0\rangle$ and displacing the spin at site $j$ ($\ge i$) to site $j+1$ along an infinite-long chain, while leaving the spin at site $j<i$ unchanged, a vacancy (hole) is then created at site $i$. The corresponding single-hole state reads
\begin{align}
  \label{eq:wf1}
  |\Psi\rangle_{t{\text-}J}  = \sum_i \varphi_h(i)\, \hat{T}_{i}\, |\phi_0\rangle,
\end{align}
where $\hat{T}_{i}$ denotes an operation translating all the spins at sites $j\ge i$ by one lattice constant to $j+1$ along the 1D chain direction. A vacancy or hole is thus inserted at site $i$, and the hole wave function $\varphi_h(i)$ can be regarded as the only variational parameter once the half-filling ground state $|\phi_0\rangle$ is known. In Appendix~\ref{Appen:sign_wf}, we give an explicit check that the wave function Eq.~(\ref{eq:wf1}) satisfies the sign structure requirement of the $t$-$J$ model. The hole wave function $\varphi_h(i)$ is also given by minimizing the hopping energy.

In terms of Eqs.~(\ref{eq:Marshall_basis}) and (\ref{eq:hfgs}), one has $\hat{T}_{i}\, |\phi_0\rangle=\sum_{\{s\}} c\left(\{s\}\right) \hat{T}_{i}\,|\{s\}\rangle$, with
\begin{align}
  \label{eq:wf2}
\hat{T}_{i}\,|\{s\}\rangle \rightarrow \E^{-\I\hat{\Omega}_i} |i; \{s\}\rangle_{ss}~.
\end{align}
Here in the newly defined Marshall basis $|i; \{s\}\rangle_{ss}$ with a hole at site $i$, the Marshall sign in Eq.~(\ref{eq:Marshall_basis}) is replaced by $ (-1)^{N_{\bar A}^\downarrow} $ on the displaced spin lattice excluding the site $i$, and Eq.~(\ref{eq:wf2}) is obtained by noting
\begin{align}
  \label{eq:signPsi}\nonumber
  (-1)^{N_{ A}^\downarrow} &= (-1)^{N_{L(i)\cap \bar A}^\downarrow + N_{R(i)\cap \bar B}^\downarrow} \\
  &=  (-1)^{N_{\bar A}^\downarrow+N_{R(i)}^\downarrow }\nonumber \\
  &\equiv  (-1)^{N_{\bar A}^\downarrow } \E^{-\I\hat{\Omega}_i} ,
\end{align}
where $L(i)$ [$R(i)$] denotes the sites on the left (right) hand side of the site $i$, and $\bar A$ and $\bar B$ are the two sublattices of the new lattice. With the hole inserted at site $i$, the sublattices $A$ and $B$ are switched for all the sites at $j>i$ in the new (the so-called squeezed) spin chain \cite{Ogata1990,Weng1991,Weng1992,Shiba1992,Weng1994,Kruis2004}.

Comparing Eqs.~(\ref{eq:vwf0}) with Eq.~(\ref{eq:wf1}), one finds the correspondence
\begin{align}
  \label{eq:vwf2}
\hat{P}_{i}\, c_{i\sigma}\,  |\phi_0\rangle \leftrightarrow (-\sigma)^i\sum_{\{s\}} c\left(\{s\}\right) |i; \{s\}\rangle_{ss},
\end{align}
where a sign factor $(-\sigma)^i$ arises from the annihilation of the spin $\sigma$ by  $c_{i\sigma}$ on the right hand side, which has already been seen in the exact expression Eq.~(\ref{eq:tJ_GSsign}). Therefore, the effect created by $ \hat{P}_{i}$, even though sign-free, is still important in 1D due to the gapless spin excitations. In the next, we outline the VMC procedure based on such variational wave functions.\\

\subsection{Variational Monte Carlo calculation}
\label{sec:MC}

The variational wave functions Eqs.~(\ref{eq:wf1}) and (\ref{eq:wf2}) can be fully constructed if we know the ground state $|\phi_0\rangle$ of the half-filled Heisenberg spin model. The half-filled state $|\phi_0\rangle$ can be approximated by the Liang-Doucot-Anderson type resonating-valence-bond (RVB) state as \cite{LDA1988}
\begin{align}
  \label{eq:RVB}
  |\phi_0\rangle &= \sum_{v} \left(\prod_{(ij)\in v} h_{ij}\right) |v\rangle,
\end{align}
where the valence bond state is given by
\begin{align}
  \label{}
  |v\rangle &= \sum_{\{\sigma\}} \left(\prod_{(ij)\in v} \epsilon_{\sigma_i,\sigma_j}\right) c_{1\sigma_1}^\dagger ... c_{L\sigma_L}^\dagger |0\rangle.
\end{align}
The Levi-Civita symbol $\epsilon_{\sigma_i,\sigma_j}$ ensures the singlet paring between spins on sites $i$ and $j$. $h_{ij}$'s are \emph{non-negative} variational parameters depending on sites $i$ and $j$ belonging to opposite sublattices, respectively \cite{LDA1988}. The most essential property of the RVB state Eq.~(\ref{eq:RVB}) is that it satisfies the exact Marshall sign rule \cite{Marshall1955} for bipartite Heisenberg models.

Based on the RVB approximation of the half-filled ground state $|\phi_0\rangle$ in Eq.~(\ref{eq:RVB}), we calculate the physical properties of the single-hole-doped variational wave functions Eqs.~(\ref{eq:wf1}) and (\ref{eq:wf2}) using the Monte Carlo method \cite{LDA1988,Sandvik2010}. The details of the VMC simulations are present in Appendix~\ref{Appen:MC}.

The variational ground state energy of the wave function Eq.~(\ref{eq:wf1}) is shown in Fig~\ref{Fig:E_compare}, which is in excellent agreement with that obtained by DMRG. By the VMC, we also calculate the momentum distributions $n_{k\alpha}$ of the variational ground states, Eqs.~(\ref{eq:wf1}) and (\ref{eq:wf2}) for the the $t$-$J$ and $\sigma$$\cdot$$t$-$J$ models, respectively. The results are presented in Fig.~\ref{Fig:MC}. The overall shapes of the curves, in particular the dip positions, are in good agreements with the theoretical predictions in Table~\ref{tab:peaks} and the DMRG results in Fig.~\ref{Fig:1-nk} . The peak values of $1-n_{k\uparrow}-n_{k\downarrow}$ are also scaled in a power law fashion as $\sim L^{-\alpha}$ for both models in Fig.~\ref{Fig:MC}(c), which are slightly different from the DMRG results possibly due to the reason that the squeezed spin chain approximation is only accurate at $t\gg J$, but here we considered $t/J=3$. Furthermore, the variational wave functions constructed here are in the sector $S^z_\mathrm{tot}=0$ based on the squeezed chain approximation, whereas the DMRG are calculated in the sector $S^z_\mathrm{tot}=\pm1/2$. This may also explain why $n_{k\uparrow}$ and $n_{k\downarrow}$ are precisely the same in Fig.~\ref{Fig:MC}(a) in contrast to a small relative shift of the DMRG in Fig.~\ref{Fig:1-nk}(a).

\section{Discussion}

The behavior of a single hole doped into a 1D Heisenberg spin chain has been examined by combined analytic, numerical, and variational approaches in the present paper. This is one of the simplest limits of doped Mott systems, involving only a single hole interacting with an antiferromagnetically correlated spin background. The basic message, not surprisingly, is that the doped hole does not propagate like a Bloch-wave due to strong correlation.

In particular, we have established a general connection of the Mott strong correlation with a specific form of many-body phase shift. That is, the hole gains a ``scattering'' phase shift $\pi$ by passing by (exchanged with) each down spin, and therefore accumulates a many-body phase string $\tau_c$ along an arbitrary path $c$ [cf. Eq.~(\ref{tauc})]. If the 1D chain is open on the two ends, we have further proved that $\tau_c$ can be explicitly incorporated into the ground state wave function by a nonlocal sign structure $\E^{-\I\hat{\Omega}_i}$ [cf. Eqs.~(\ref{eq:tJ_GSsign}) and (\ref{eq:ps})].

In a conventional many-body fermion system, the ground state wave function satisfies the fermion sign structure, which dictates the momentum structure with a Fermi surface satisfying the Luttinger volume and a finite $Z_k$ at the Fermi surface. By contrast, the fermion sign structure gets completely altered \cite{zaanen2011mottness} by strong on-site Coulomb repulsion and the new sign structure, i.e., $\tau_c$ and $\E^{-\I\hat{\Omega}_i}$ here, determines the momentum distribution of a non-Fermi-liquid or Luttinger liquid behavior with vanishing $Z_k$ in the large $L$ limit.

As it turns out, the chief role of $\tau_c$ is not a mass renormalization. As a matter of fact, the effective mass is even not affected by $\tau_c$ in the case of an open chain. However, the momentum structure is completely decided by it. The profile of the momentum distribution shown in Fig.~ \ref{Fig:1-nk} is categorically different from a residual Fermi distribution, even though the Luttinger volume seems unchanged as compared to a non-interacting electrons of the same density (which does not distinguish half-filling and one-hole-doping in the thermodynamic limit). In fact, the momentum distribution is completely flat at half-filling due to the strong correlation. The emergent quasiparticle spectral weight $Z_{k_0}$ of the hole is found to vanish as $\approx L^{-0.49}$ at large chain length $L$  (at $t/J=3$) by DMRG. For the periodic boundary condition, the interference effect of $\tau_c$ has been also clearly identified.

These results are in sharp contrast with those obtained by switching off the phase string sign structure. Indeed, in the $\sigma$$\cdot$$t$-$J$ model without $\tau_c$, the characteristic momentum of the hole at $k_0=\pm \pi/2$ is then shifted to $\pi$, and  $Z_{k_0}$ follows a different scaling law, vanishing slower $\approx L^{-0.23}$ at large $L$. In other words, the sign structure of the $t$-$J$ model does play a critical role in shaping the motion of the hole on a quantum spin background and its singular effect must be treated with a great care. This has been further confirmed by the variational wave function approach based on the $t$-$J$ and $\sigma$$\cdot$$t$-$J$ models via VMC in this work.

Since the sign structure $\tau_c$ has been precisely identified in the  $t$-$J$ model for any dimensions \cite{Sheng1996,Weng1997,Wu2008sign}, one expects that the novel properties of doped Mott insulators, due to the irreparable many-body phase shift, should persist beyond the 1D case as well as beyond the single-hole-doping case \cite{zaanen2011mottness}.

For example, recently a single hole doped into a two-leg Heisenberg spin ladder has been studied by both the DMRG \cite{ZZ2013,ZZ2014,ZZ2014qp,ZZ2014cm,Kivelson} and a wave function approach \cite{ladder2015} based on the VMC. In this system, the ``vacuum'' of the spin background is spin gapped at half-filling. So only a finite-size ``cloud'' of spin excitations can be created around the doped hole. In the strong anisotropic or strong rung limit, such a spin cloud or spin polaron effect is indeed found only to renormalize the effective mass without changing the Bloch-wave nature. But with reducing the anisotropy, a critical point can be reached \cite{ZZ2014qp,ZZ2014cm}, beyond which the momentum structure is fundamentally changed accompanied by the charge modulation and the divergence of the charge mass \cite{ZZ2013,ZZ2014qp}. It has been demonstrated that the singular many-body phase shift or phase string $\tau_c$ becomes unscreened here and is responsible for these exotic properties \cite{ZZ2013,ZZ2014qp,ZZ2014cm,ladder2015}.

Such a novel phase of the two-ladder system is shown to smoothly persist in the limit of strong-chain/weak-rung coupling. In the extreme limit of vanishing rung coupling, the two-leg ladder further reduces to two decoupled 1D $t$-$J$ chains, with the hole localized within one of the chains. This corresponds to the single-hole-doped Heisenberg spin chain studied in the present paper. As compared to the coupled two-leg ladder, there are two basic distinctions in this limit. One is that the spin background is now gapless in 1D, instead of spin-gapped in the coupled two-leg case at half-filling. In other words, the spin-polaron effect becomes more important in 1D, in a form of spin-charge separation even for the $\sigma$$\cdot$$t$-$J$ model. The second distinction is that in the two-leg ladder there is an important quantum interference effect originated from the hole transversing from different paths \cite{ZZ2013}, but it is obviously absent in the 1D case. Of course, as shown in this paper, the nontrivial quantum interference still takes place as a finite-size effect under the periodic boundary condition, where the hole may reach a lattice site either through a shorter path or by circumventing the closed ring.

Therefore, the 1D non-Fermi-liquid behavior, the reconstruction of momentum structure in both 1D and the ladder systems, the charge modulation and possible self-localization \cite{ZZ2013,ZZ2014qp,ZZ2014cm} in a ladder system with the leg number more than one, as well as the strong pairing mechanism in the even-leg ladder systems \cite{ZZ2014}, have so far all been attributed to the exotic sign structure $\tau_c$ originated from the Mott physics. We expect the same sign structure to play a critical role in a 2D system as well, which might be directly relevant to the high-$T_c$ cuprate.

\begin{acknowledgements}
Useful discussion with J. Zaanen is acknowledged. This work is supported by the NBRC (973 Program, Nos. 2015CB921000), NSFC Grant no. 11534007 and US National Science Foundation Grant  DMR-1408560.
\end{acknowledgements}


\begin{thebibliography}{41}%
\makeatletter
\providecommand \@ifxundefined [1]{%
 \@ifx{#1\undefined}
}%
\providecommand \@ifnum [1]{%
 \ifnum #1\expandafter \@firstoftwo
 \else \expandafter \@secondoftwo
 \fi
}%
\providecommand \@ifx [1]{%
 \ifx #1\expandafter \@firstoftwo
 \else \expandafter \@secondoftwo
 \fi
}%
\providecommand \natexlab [1]{#1}%
\providecommand \enquote  [1]{``#1''}%
\providecommand \bibnamefont  [1]{#1}%
\providecommand \bibfnamefont [1]{#1}%
\providecommand \citenamefont [1]{#1}%
\providecommand \href@noop [0]{\@secondoftwo}%
\providecommand \href [0]{\begingroup \@sanitize@url \@href}%
\providecommand \@href[1]{\@@startlink{#1}\@@href}%
\providecommand \@@href[1]{\endgroup#1\@@endlink}%
\providecommand \@sanitize@url [0]{\catcode `\\12\catcode `\$12\catcode
  `\&12\catcode `\#12\catcode `\^12\catcode `\_12\catcode `\%12\relax}%
\providecommand \@@startlink[1]{}%
\providecommand \@@endlink[0]{}%
\providecommand \url  [0]{\begingroup\@sanitize@url \@url }%
\providecommand \@url [1]{\endgroup\@href {#1}{\urlprefix }}%
\providecommand \urlprefix  [0]{URL }%
\providecommand \Eprint [0]{\href }%
\providecommand \doibase [0]{http://dx.doi.org/}%
\providecommand \selectlanguage [0]{\@gobble}%
\providecommand \bibinfo  [0]{\@secondoftwo}%
\providecommand \bibfield  [0]{\@secondoftwo}%
\providecommand \translation [1]{[#1]}%
\providecommand \BibitemOpen [0]{}%
\providecommand \bibitemStop [0]{}%
\providecommand \bibitemNoStop [0]{.\EOS\space}%
\providecommand \EOS [0]{\spacefactor3000\relax}%
\providecommand \BibitemShut  [1]{\csname bibitem#1\endcsname}%
\let\auto@bib@innerbib\@empty
\bibitem [{\citenamefont {Anderson}(1997)}]{Anderson}%
  \BibitemOpen
  \bibfield  {author} {\bibinfo {author} {\bibfnamefont {P.~W.}\ \bibnamefont
  {Anderson}},\ }\href
  {http://www.amazon.com/Theory-Superconductivity-High-Tc-Cuprate-Superconductors/dp/0691043655}
  {\emph {\bibinfo {title} {{The Theory of Superconductivity in the High-Tc
  Cuprate Superconductors}}}}\ (\bibinfo  {publisher} {Princeton University
  Press, Princeton, NJ},\ \bibinfo {year} {1997})\BibitemShut {NoStop}%
\bibitem [{\citenamefont {Lee}\ \emph {et~al.}(2006)\citenamefont {Lee},
  \citenamefont {Nagaosa},\ and\ \citenamefont {Wen}}]{Lee2006}%
  \BibitemOpen
  \bibfield  {author} {\bibinfo {author} {\bibfnamefont {P.~A.}\ \bibnamefont
  {Lee}}, \bibinfo {author} {\bibfnamefont {N.}~\bibnamefont {Nagaosa}}, \ and\
  \bibinfo {author} {\bibfnamefont {X.-G.}\ \bibnamefont {Wen}},\ }\href
  {\doibase 10.1103/RevModPhys.78.17} {\bibfield  {journal} {\bibinfo
  {journal} {Rev. Mod. Phys.}\ }\textbf {\bibinfo {volume} {78}},\ \bibinfo
  {pages} {17} (\bibinfo {year} {2006})}\BibitemShut {NoStop}%
\bibitem [{\citenamefont {Anderson}(1987)}]{Anderson1987}%
  \BibitemOpen
  \bibfield  {author} {\bibinfo {author} {\bibfnamefont {P.~W.}\ \bibnamefont
  {Anderson}},\ }\href {\doibase 10.1126/science.235.4793.1196} {\bibfield
  {journal} {\bibinfo  {journal} {Science}\ }\textbf {\bibinfo {volume}
  {235}},\ \bibinfo {pages} {1196} (\bibinfo {year} {1987})}\BibitemShut
  {NoStop}%
\bibitem [{\citenamefont {Lieb}\ and\ \citenamefont {Wu}(1968)}]{Lieb1968}%
  \BibitemOpen
  \bibfield  {author} {\bibinfo {author} {\bibfnamefont {E.~H.}\ \bibnamefont
  {Lieb}}\ and\ \bibinfo {author} {\bibfnamefont {F.~Y.}\ \bibnamefont {Wu}},\
  }\href {\doibase 10.1103/PhysRevLett.20.1445} {\bibfield  {journal} {\bibinfo
   {journal} {Phys. Rev. Lett.}\ }\textbf {\bibinfo {volume} {20}},\ \bibinfo
  {pages} {1445} (\bibinfo {year} {1968})}\BibitemShut {NoStop}%
\bibitem [{\citenamefont {Woynarovich}(1982)}]{Woynarovich1982}%
  \BibitemOpen
  \bibfield  {author} {\bibinfo {author} {\bibfnamefont {F.}~\bibnamefont
  {Woynarovich}},\ }\href {http://stacks.iop.org/0022-3719/15/i=1/a=007}
  {\bibfield  {journal} {\bibinfo  {journal} {J. Phys. C Solid State Phys.}\
  }\textbf {\bibinfo {volume} {15}},\ \bibinfo {pages} {85} (\bibinfo {year}
  {1982})}\BibitemShut {NoStop}%
\bibitem [{\citenamefont {Ogata}\ and\ \citenamefont
  {Shiba}(1990)}]{Ogata1990}%
  \BibitemOpen
  \bibfield  {author} {\bibinfo {author} {\bibfnamefont {M.}~\bibnamefont
  {Ogata}}\ and\ \bibinfo {author} {\bibfnamefont {H.}~\bibnamefont {Shiba}},\
  }\href {\doibase 10.1103/PhysRevB.41.2326} {\bibfield  {journal} {\bibinfo
  {journal} {Phys. Rev. B}\ }\textbf {\bibinfo {volume} {41}},\ \bibinfo
  {pages} {2326} (\bibinfo {year} {1990})}\BibitemShut {NoStop}%
\bibitem [{\citenamefont {Ogata}\ \emph
  {et~al.}(1991{\natexlab{a}})\citenamefont {Ogata}, \citenamefont {Sugiyama},\
  and\ \citenamefont {Shiba}}]{Ogata1991b}%
  \BibitemOpen
  \bibfield  {author} {\bibinfo {author} {\bibfnamefont {M.}~\bibnamefont
  {Ogata}}, \bibinfo {author} {\bibfnamefont {T.}~\bibnamefont {Sugiyama}}, \
  and\ \bibinfo {author} {\bibfnamefont {H.}~\bibnamefont {Shiba}},\ }\href
  {\doibase 10.1103/PhysRevB.43.8401} {\bibfield  {journal} {\bibinfo
  {journal} {Phys. Rev. B}\ }\textbf {\bibinfo {volume} {43}},\ \bibinfo
  {pages} {8401} (\bibinfo {year} {1991}{\natexlab{a}})}\BibitemShut {NoStop}%
\bibitem [{\citenamefont {Sutherland}(1975)}]{Sutherland1975}%
  \BibitemOpen
  \bibfield  {author} {\bibinfo {author} {\bibfnamefont {B.}~\bibnamefont
  {Sutherland}},\ }\href {\doibase 10.1103/PhysRevB.12.3795} {\bibfield
  {journal} {\bibinfo  {journal} {Phys. Rev. B}\ }\textbf {\bibinfo {volume}
  {12}},\ \bibinfo {pages} {3795} (\bibinfo {year} {1975})}\BibitemShut
  {NoStop}%
\bibitem [{\citenamefont {Schlottmann}(1987)}]{Schlottmann1987}%
  \BibitemOpen
  \bibfield  {author} {\bibinfo {author} {\bibfnamefont {P.}~\bibnamefont
  {Schlottmann}},\ }\href {\doibase 10.1103/PhysRevB.36.5177} {\bibfield
  {journal} {\bibinfo  {journal} {Phys. Rev. B}\ }\textbf {\bibinfo {volume}
  {36}},\ \bibinfo {pages} {5177} (\bibinfo {year} {1987})}\BibitemShut
  {NoStop}%
\bibitem [{\citenamefont {Bares}\ and\ \citenamefont
  {Blatter}(1990)}]{Bares1990}%
  \BibitemOpen
  \bibfield  {author} {\bibinfo {author} {\bibfnamefont {P.~A.}\ \bibnamefont
  {Bares}}\ and\ \bibinfo {author} {\bibfnamefont {G.}~\bibnamefont
  {Blatter}},\ }\href {\doibase 10.1103/PhysRevLett.64.2567} {\bibfield
  {journal} {\bibinfo  {journal} {Phys. Rev. Lett.}\ }\textbf {\bibinfo
  {volume} {64}},\ \bibinfo {pages} {2567} (\bibinfo {year}
  {1990})}\BibitemShut {NoStop}%
\bibitem [{\citenamefont {Bares}\ \emph {et~al.}(1991)\citenamefont {Bares},
  \citenamefont {Blatter},\ and\ \citenamefont {Ogata}}]{Bares1991}%
  \BibitemOpen
  \bibfield  {author} {\bibinfo {author} {\bibfnamefont {P.~A.}\ \bibnamefont
  {Bares}}, \bibinfo {author} {\bibfnamefont {G.}~\bibnamefont {Blatter}}, \
  and\ \bibinfo {author} {\bibfnamefont {M.}~\bibnamefont {Ogata}},\ }\href
  {\doibase 10.1103/PhysRevB.44.130} {\bibfield  {journal} {\bibinfo  {journal}
  {Phys. Rev. B}\ }\textbf {\bibinfo {volume} {44}},\ \bibinfo {pages} {130}
  (\bibinfo {year} {1991})}\BibitemShut {NoStop}%
\bibitem [{\citenamefont {Haldane}(1981{\natexlab{a}})}]{Haldane1981a}%
  \BibitemOpen
  \bibfield  {author} {\bibinfo {author} {\bibfnamefont {F.~D.~M.}\
  \bibnamefont {Haldane}},\ }\href@noop {} {\bibfield  {journal} {\bibinfo
  {journal} {J. Phys. C Solid State Phys.}\ }\textbf {\bibinfo {volume} {14}},\
  \bibinfo {pages} {2585} (\bibinfo {year} {1981}{\natexlab{a}})}\BibitemShut
  {NoStop}%
\bibitem [{\citenamefont {Haldane}(1981{\natexlab{b}})}]{Haldane1981b}%
  \BibitemOpen
  \bibfield  {author} {\bibinfo {author} {\bibfnamefont {F.~D.~M.}\
  \bibnamefont {Haldane}},\ }\href {\doibase 10.1103/PhysRevLett.47.1840}
  {\bibfield  {journal} {\bibinfo  {journal} {Phys. Rev. Lett.}\ }\textbf
  {\bibinfo {volume} {47}},\ \bibinfo {pages} {1840} (\bibinfo {year}
  {1981}{\natexlab{b}})}\BibitemShut {NoStop}%
\bibitem [{\citenamefont {Ogata}\ \emph
  {et~al.}(1991{\natexlab{b}})\citenamefont {Ogata}, \citenamefont {Luchini},
  \citenamefont {Sorella},\ and\ \citenamefont {Assaad}}]{Ogata1991}%
  \BibitemOpen
  \bibfield  {author} {\bibinfo {author} {\bibfnamefont {M.}~\bibnamefont
  {Ogata}}, \bibinfo {author} {\bibfnamefont {M.}~\bibnamefont {Luchini}},
  \bibinfo {author} {\bibfnamefont {S.}~\bibnamefont {Sorella}}, \ and\
  \bibinfo {author} {\bibfnamefont {F.}~\bibnamefont {Assaad}},\ }\href
  {\doibase 10.1103/PhysRevLett.66.2388} {\bibfield  {journal} {\bibinfo
  {journal} {Phys. Rev. Lett.}\ }\textbf {\bibinfo {volume} {66}},\ \bibinfo
  {pages} {2388} (\bibinfo {year} {1991}{\natexlab{b}})}\BibitemShut {NoStop}%
\bibitem [{\citenamefont {Dagotto}(1994)}]{Dagotto1994}%
  \BibitemOpen
  \bibfield  {author} {\bibinfo {author} {\bibfnamefont {E.}~\bibnamefont
  {Dagotto}},\ }\href {\doibase 10.1103/RevModPhys.66.763} {\bibfield
  {journal} {\bibinfo  {journal} {Rev. Mod. Phys.}\ }\textbf {\bibinfo {volume}
  {66}},\ \bibinfo {pages} {763} (\bibinfo {year} {1994})}\BibitemShut
  {NoStop}%
\bibitem [{\citenamefont {Moreno}\ \emph {et~al.}(2011)\citenamefont {Moreno},
  \citenamefont {Muramatsu},\ and\ \citenamefont {Manmana}}]{Moreno2011}%
  \BibitemOpen
  \bibfield  {author} {\bibinfo {author} {\bibfnamefont {A.}~\bibnamefont
  {Moreno}}, \bibinfo {author} {\bibfnamefont {A.}~\bibnamefont {Muramatsu}}, \
  and\ \bibinfo {author} {\bibfnamefont {S.~R.}\ \bibnamefont {Manmana}},\
  }\href {\doibase 10.1103/PhysRevB.83.205113} {\bibfield  {journal} {\bibinfo
  {journal} {Phys. Rev. B}\ }\textbf {\bibinfo {volume} {83}},\ \bibinfo
  {pages} {205113} (\bibinfo {year} {2011})}\BibitemShut {NoStop}%
\bibitem [{\citenamefont {Anderson}(1990)}]{Anderson90PRL}%
  \BibitemOpen
  \bibfield  {author} {\bibinfo {author} {\bibfnamefont {P.}~\bibnamefont
  {Anderson}},\ }\href {\doibase 10.1103/PhysRevLett.64.1839} {\bibfield
  {journal} {\bibinfo  {journal} {Phys. Rev. Lett.}\ }\textbf {\bibinfo
  {volume} {64}},\ \bibinfo {pages} {1839} (\bibinfo {year}
  {1990})}\BibitemShut {NoStop}%
\bibitem [{\citenamefont {Ren}\ and\ \citenamefont
  {Anderson}(1993)}]{Ren-Anderson1993}%
  \BibitemOpen
  \bibfield  {author} {\bibinfo {author} {\bibfnamefont {Y.}~\bibnamefont
  {Ren}}\ and\ \bibinfo {author} {\bibfnamefont {P.~W.}\ \bibnamefont
  {Anderson}},\ }\href {\doibase 10.1103/PhysRevB.48.16662} {\bibfield
  {journal} {\bibinfo  {journal} {Phys. Rev. B}\ }\textbf {\bibinfo {volume}
  {48}},\ \bibinfo {pages} {16662} (\bibinfo {year} {1993})}\BibitemShut
  {NoStop}%
\bibitem [{\citenamefont {Weng}\ \emph {et~al.}(1991)\citenamefont {Weng},
  \citenamefont {Sheng}, \citenamefont {Ting},\ and\ \citenamefont
  {Su}}]{Weng1991}%
  \BibitemOpen
  \bibfield  {author} {\bibinfo {author} {\bibfnamefont {Z.~Y.}\ \bibnamefont
  {Weng}}, \bibinfo {author} {\bibfnamefont {D.~N.}\ \bibnamefont {Sheng}},
  \bibinfo {author} {\bibfnamefont {C.~S.}\ \bibnamefont {Ting}}, \ and\
  \bibinfo {author} {\bibfnamefont {Z.~B.}\ \bibnamefont {Su}},\ }\href
  {\doibase 10.1103/PhysRevLett.67.3318} {\bibfield  {journal} {\bibinfo
  {journal} {Phys. Rev. Lett.}\ }\textbf {\bibinfo {volume} {67}},\ \bibinfo
  {pages} {3318} (\bibinfo {year} {1991})}\BibitemShut {NoStop}%
\bibitem [{\citenamefont {Weng}\ \emph {et~al.}(1992)\citenamefont {Weng},
  \citenamefont {Sheng}, \citenamefont {Ting},\ and\ \citenamefont
  {Su}}]{Weng1992}%
  \BibitemOpen
  \bibfield  {author} {\bibinfo {author} {\bibfnamefont {Z.~Y.}\ \bibnamefont
  {Weng}}, \bibinfo {author} {\bibfnamefont {D.~N.}\ \bibnamefont {Sheng}},
  \bibinfo {author} {\bibfnamefont {C.~S.}\ \bibnamefont {Ting}}, \ and\
  \bibinfo {author} {\bibfnamefont {Z.~B.}\ \bibnamefont {Su}},\ }\href
  {\doibase 10.1103/PhysRevB.45.7850} {\bibfield  {journal} {\bibinfo
  {journal} {Phys. Rev. B}\ }\textbf {\bibinfo {volume} {45}},\ \bibinfo
  {pages} {7850} (\bibinfo {year} {1992})}\BibitemShut {NoStop}%
\bibitem [{\citenamefont {Wang}\ and\ \citenamefont {Ye}(2014)}]{Wang2014}%
  \BibitemOpen
  \bibfield  {author} {\bibinfo {author} {\bibfnamefont {Q.-R.}\ \bibnamefont
  {Wang}}\ and\ \bibinfo {author} {\bibfnamefont {P.}~\bibnamefont {Ye}},\
  }\href {\doibase 10.1103/PhysRevB.90.045106} {\bibfield  {journal} {\bibinfo
  {journal} {Phys. Rev. B}\ }\textbf {\bibinfo {volume} {90}},\ \bibinfo
  {pages} {45106} (\bibinfo {year} {2014})}\BibitemShut {NoStop}%
\bibitem [{\citenamefont {Shiba}\ and\ \citenamefont
  {Ogata}(1992)}]{Shiba1992}%
  \BibitemOpen
  \bibfield  {author} {\bibinfo {author} {\bibfnamefont {H.}~\bibnamefont
  {Shiba}}\ and\ \bibinfo {author} {\bibfnamefont {M.}~\bibnamefont {Ogata}},\
  }\href {\doibase 10.1143/PTPS.108.265} {\bibfield  {journal} {\bibinfo
  {journal} {Prog. Theor. Phys. Suppl.}\ }\textbf {\bibinfo {volume} {108}},\
  \bibinfo {pages} {265} (\bibinfo {year} {1992})}\BibitemShut {NoStop}%
\bibitem [{\citenamefont {Weng}(1994)}]{Weng1994}%
  \BibitemOpen
  \bibfield  {author} {\bibinfo {author} {\bibfnamefont {Z.~Y.}\ \bibnamefont
  {Weng}},\ }\href {\doibase 10.1103/PhysRevB.50.13837} {\bibfield  {journal}
  {\bibinfo  {journal} {Phys. Rev. B}\ }\textbf {\bibinfo {volume} {50}},\
  \bibinfo {pages} {13837} (\bibinfo {year} {1994})}\BibitemShut {NoStop}%
\bibitem [{\citenamefont {Weng}\ \emph {et~al.}(1997)\citenamefont {Weng},
  \citenamefont {Sheng}, \citenamefont {Chen},\ and\ \citenamefont
  {Ting}}]{Weng1997}%
  \BibitemOpen
  \bibfield  {author} {\bibinfo {author} {\bibfnamefont {Z.~Y.}\ \bibnamefont
  {Weng}}, \bibinfo {author} {\bibfnamefont {D.~N.}\ \bibnamefont {Sheng}},
  \bibinfo {author} {\bibfnamefont {Y.~C.}\ \bibnamefont {Chen}}, \ and\
  \bibinfo {author} {\bibfnamefont {C.~S.}\ \bibnamefont {Ting}},\ }\href
  {\doibase 10.1103/PhysRevB.55.3894} {\bibfield  {journal} {\bibinfo
  {journal} {Phys. Rev. B}\ }\textbf {\bibinfo {volume} {55}},\ \bibinfo
  {pages} {3894} (\bibinfo {year} {1997})}\BibitemShut {NoStop}%
\bibitem [{\citenamefont {Kruis}\ \emph {et~al.}(2004)\citenamefont {Kruis},
  \citenamefont {McCulloch}, \citenamefont {Nussinov},\ and\ \citenamefont
  {Zaanen}}]{Kruis2004}%
  \BibitemOpen
  \bibfield  {author} {\bibinfo {author} {\bibfnamefont {H.~V.}\ \bibnamefont
  {Kruis}}, \bibinfo {author} {\bibfnamefont {I.~P.}\ \bibnamefont
  {McCulloch}}, \bibinfo {author} {\bibfnamefont {Z.}~\bibnamefont {Nussinov}},
  \ and\ \bibinfo {author} {\bibfnamefont {J.}~\bibnamefont {Zaanen}},\ }\href
  {\doibase 10.1103/PhysRevB.70.075109} {\bibfield  {journal} {\bibinfo
  {journal} {Phys. Rev. B}\ }\textbf {\bibinfo {volume} {70}},\ \bibinfo
  {pages} {075109} (\bibinfo {year} {2004})}\BibitemShut {NoStop}%
\bibitem [{\citenamefont {Zhu}\ \emph {et~al.}(2013)\citenamefont {Zhu},
  \citenamefont {Jiang}, \citenamefont {Qi}, \citenamefont {Tian},\ and\
  \citenamefont {Weng}}]{ZZ2013}%
  \BibitemOpen
  \bibfield  {author} {\bibinfo {author} {\bibfnamefont {Z.}~\bibnamefont
  {Zhu}}, \bibinfo {author} {\bibfnamefont {H.-C.}\ \bibnamefont {Jiang}},
  \bibinfo {author} {\bibfnamefont {Y.}~\bibnamefont {Qi}}, \bibinfo {author}
  {\bibfnamefont {C.-S.}\ \bibnamefont {Tian}}, \ and\ \bibinfo {author}
  {\bibfnamefont {Z.-Y.}\ \bibnamefont {Weng}},\ }\href {\doibase
  10.1038/srep02586} {\bibfield  {journal} {\bibinfo  {journal} {Sci. Rep.}\
  }\textbf {\bibinfo {volume} {3}},\ \bibinfo {pages} {2586} (\bibinfo {year}
  {2013})}\BibitemShut {NoStop}%
\bibitem [{\citenamefont {Sheng}\ \emph {et~al.}(1996)\citenamefont {Sheng},
  \citenamefont {Chen},\ and\ \citenamefont {Weng}}]{Sheng1996}%
  \BibitemOpen
  \bibfield  {author} {\bibinfo {author} {\bibfnamefont {D.~N.}\ \bibnamefont
  {Sheng}}, \bibinfo {author} {\bibfnamefont {Y.~C.}\ \bibnamefont {Chen}}, \
  and\ \bibinfo {author} {\bibfnamefont {Z.~Y.}\ \bibnamefont {Weng}},\ }\href
  {\doibase 10.1103/PhysRevLett.77.5102} {\bibfield  {journal} {\bibinfo
  {journal} {Phys. Rev. Lett.}\ }\textbf {\bibinfo {volume} {77}},\ \bibinfo
  {pages} {5102} (\bibinfo {year} {1996})}\BibitemShut {NoStop}%
\bibitem [{\citenamefont {Wu}\ \emph {et~al.}(2008)\citenamefont {Wu},
  \citenamefont {Weng},\ and\ \citenamefont {Zaanen}}]{Wu2008sign}%
  \BibitemOpen
  \bibfield  {author} {\bibinfo {author} {\bibfnamefont {K.}~\bibnamefont
  {Wu}}, \bibinfo {author} {\bibfnamefont {Z.~Y.}\ \bibnamefont {Weng}}, \ and\
  \bibinfo {author} {\bibfnamefont {J.}~\bibnamefont {Zaanen}},\ }\href
  {\doibase 10.1103/PhysRevB.77.155102} {\bibfield  {journal} {\bibinfo
  {journal} {Phys. Rev. B}\ }\textbf {\bibinfo {volume} {77}},\ \bibinfo
  {pages} {155102} (\bibinfo {year} {2008})}\BibitemShut {NoStop}%
\bibitem [{\citenamefont {Marshall}(1955)}]{Marshall1955}%
  \BibitemOpen
  \bibfield  {author} {\bibinfo {author} {\bibfnamefont {W.}~\bibnamefont
  {Marshall}},\ }\href {\doibase 10.1098/rspa.1955.0200} {\bibfield  {journal}
  {\bibinfo  {journal} {Proc. R. Soc. Lond. A}\ }\textbf {\bibinfo {volume}
  {232}},\ \bibinfo {pages} {48} (\bibinfo {year} {1955})}\BibitemShut
  {NoStop}%
\bibitem [{\citenamefont {Poilblanc}(2008)}]{Poilblanc2008}%
  \BibitemOpen
  \bibfield  {author} {\bibinfo {author} {\bibfnamefont {D.}~\bibnamefont
  {Poilblanc}},\ }\href {\doibase 10.1103/PhysRevLett.100.157206} {\bibfield
  {journal} {\bibinfo  {journal} {Phys. Rev. Lett.}\ }\textbf {\bibinfo
  {volume} {100}},\ \bibinfo {pages} {157206} (\bibinfo {year}
  {2008})}\BibitemShut {NoStop}%
\bibitem [{\citenamefont {Suzuura}\ and\ \citenamefont
  {Nagaosa}(1997)}]{Suzuura1997}%
  \BibitemOpen
  \bibfield  {author} {\bibinfo {author} {\bibfnamefont {H.}~\bibnamefont
  {Suzuura}}\ and\ \bibinfo {author} {\bibfnamefont {N.}~\bibnamefont
  {Nagaosa}},\ }\href {\doibase 10.1103/PhysRevB.56.3548} {\bibfield  {journal}
  {\bibinfo  {journal} {Phys. Rev. B}\ }\textbf {\bibinfo {volume} {56}},\
  \bibinfo {pages} {3548} (\bibinfo {year} {1997})}\BibitemShut {NoStop}%
\bibitem [{\citenamefont {Nagaosa}(1998)}]{Nagaosa1998}%
  \BibitemOpen
  \bibfield  {author} {\bibinfo {author} {\bibfnamefont {N.}~\bibnamefont
  {Nagaosa}},\ }\href {\doibase 10.1088/0953-8984/10/49/025} {\bibfield
  {journal} {\bibinfo  {journal} {J. Phys. Condens. Matter}\ }\textbf {\bibinfo
  {volume} {10}},\ \bibinfo {pages} {11385} (\bibinfo {year}
  {1998})}\BibitemShut {NoStop}%
\bibitem [{\citenamefont {White}(1992)}]{DMRG1992}%
  \BibitemOpen
  \bibfield  {author} {\bibinfo {author} {\bibfnamefont {S.~R.}\ \bibnamefont
  {White}},\ }\href {\doibase 10.1103/PhysRevLett.69.2863} {\bibfield
  {journal} {\bibinfo  {journal} {Phys. Rev. Lett.}\ }\textbf {\bibinfo
  {volume} {69}},\ \bibinfo {pages} {2863} (\bibinfo {year}
  {1992})}\BibitemShut {NoStop}%
\bibitem [{\citenamefont {Wang}\ \emph {et~al.}(2015)\citenamefont {Wang},
  \citenamefont {Zhu}, \citenamefont {Qi},\ and\ \citenamefont
  {Weng}}]{ladder2015}%
  \BibitemOpen
  \bibfield  {author} {\bibinfo {author} {\bibfnamefont {Q.-R.}\ \bibnamefont
  {Wang}}, \bibinfo {author} {\bibfnamefont {Z.}~\bibnamefont {Zhu}}, \bibinfo
  {author} {\bibfnamefont {Y.}~\bibnamefont {Qi}}, \ and\ \bibinfo {author}
  {\bibfnamefont {Z.-Y.}\ \bibnamefont {Weng}},\ }\href@noop {} {\bibfield
  {journal} {\bibinfo  {journal} {arXiv:1509.01260}\ } (\bibinfo {year}
  {2015})}\BibitemShut {NoStop}%
\bibitem [{\citenamefont {Liang}\ \emph {et~al.}(1988)\citenamefont {Liang},
  \citenamefont {Doucot},\ and\ \citenamefont {Anderson}}]{LDA1988}%
  \BibitemOpen
  \bibfield  {author} {\bibinfo {author} {\bibfnamefont {S.}~\bibnamefont
  {Liang}}, \bibinfo {author} {\bibfnamefont {B.}~\bibnamefont {Doucot}}, \
  and\ \bibinfo {author} {\bibfnamefont {P.~W.}\ \bibnamefont {Anderson}},\
  }\href {\doibase 10.1103/PhysRevLett.61.365} {\bibfield  {journal} {\bibinfo
  {journal} {Phys. Rev. Lett.}\ }\textbf {\bibinfo {volume} {61}},\ \bibinfo
  {pages} {365} (\bibinfo {year} {1988})}\BibitemShut {NoStop}%
\bibitem [{\citenamefont {Sandvik}\ and\ \citenamefont
  {Evertz}(2010)}]{Sandvik2010}%
  \BibitemOpen
  \bibfield  {author} {\bibinfo {author} {\bibfnamefont {A.~W.}\ \bibnamefont
  {Sandvik}}\ and\ \bibinfo {author} {\bibfnamefont {H.~G.}\ \bibnamefont
  {Evertz}},\ }\href {\doibase 10.1103/PhysRevB.82.024407} {\bibfield
  {journal} {\bibinfo  {journal} {Phys. Rev. B}\ }\textbf {\bibinfo {volume}
  {82}},\ \bibinfo {pages} {24407} (\bibinfo {year} {2010})}\BibitemShut
  {NoStop}%
\bibitem [{\citenamefont {Zaanen}\ and\ \citenamefont
  {Overbosch}(2011)}]{zaanen2011mottness}%
  \BibitemOpen
  \bibfield  {author} {\bibinfo {author} {\bibfnamefont {J.}~\bibnamefont
  {Zaanen}}\ and\ \bibinfo {author} {\bibfnamefont {B.~J.}\ \bibnamefont
  {Overbosch}},\ }\href@noop {} {\bibfield  {journal} {\bibinfo  {journal}
  {Philos. Trans. R. Soc. A Math. Phys. Eng. Sci.}\ }\textbf {\bibinfo {volume}
  {369}},\ \bibinfo {pages} {1599} (\bibinfo {year} {2011})}\BibitemShut
  {NoStop}%
\bibitem [{\citenamefont {Zhu}\ \emph {et~al.}(2014)\citenamefont {Zhu},
  \citenamefont {Jiang}, \citenamefont {Sheng},\ and\ \citenamefont
  {Weng}}]{ZZ2014}%
  \BibitemOpen
  \bibfield  {author} {\bibinfo {author} {\bibfnamefont {Z.}~\bibnamefont
  {Zhu}}, \bibinfo {author} {\bibfnamefont {H.-C.}\ \bibnamefont {Jiang}},
  \bibinfo {author} {\bibfnamefont {D.~N.}\ \bibnamefont {Sheng}}, \ and\
  \bibinfo {author} {\bibfnamefont {Z.-Y.}\ \bibnamefont {Weng}},\ }\href
  {\doibase 10.1038/srep05419} {\bibfield  {journal} {\bibinfo  {journal} {Sci.
  Rep.}\ }\textbf {\bibinfo {volume} {4}},\ \bibinfo {pages} {5419} (\bibinfo
  {year} {2014})}\BibitemShut {NoStop}%
\bibitem [{\citenamefont {Zhu}\ and\ \citenamefont {Weng}(2014)}]{ZZ2014qp}%
  \BibitemOpen
  \bibfield  {author} {\bibinfo {author} {\bibfnamefont {Z.}~\bibnamefont
  {Zhu}}\ and\ \bibinfo {author} {\bibfnamefont {Z.-Y.}\ \bibnamefont {Weng}},\
  }\href@noop {} {\bibfield  {journal} {\bibinfo  {journal} {arXiv:1409.3241}\
  } (\bibinfo {year} {2014})}\BibitemShut {NoStop}%
\bibitem [{\citenamefont {Zhu}\ \emph {et~al.}(2015)\citenamefont {Zhu},
  \citenamefont {Tian}, \citenamefont {Jiang}, \citenamefont {Qi},
  \citenamefont {Weng},\ and\ \citenamefont {Zaanen}}]{ZZ2014cm}%
  \BibitemOpen
  \bibfield  {author} {\bibinfo {author} {\bibfnamefont {Z.}~\bibnamefont
  {Zhu}}, \bibinfo {author} {\bibfnamefont {C.-S.}\ \bibnamefont {Tian}},
  \bibinfo {author} {\bibfnamefont {H.-C.}\ \bibnamefont {Jiang}}, \bibinfo
  {author} {\bibfnamefont {Y.}~\bibnamefont {Qi}}, \bibinfo {author}
  {\bibfnamefont {Z.-Y.}\ \bibnamefont {Weng}}, \ and\ \bibinfo {author}
  {\bibfnamefont {J.}~\bibnamefont {Zaanen}},\ }\href {\doibase
  10.1103/PhysRevB.92.035113} {\bibfield  {journal} {\bibinfo  {journal} {Phys.
  Rev. B}\ }\textbf {\bibinfo {volume} {92}},\ \bibinfo {pages} {35113}
  (\bibinfo {year} {2015})}\BibitemShut {NoStop}%
\bibitem [{\citenamefont {White}\ \emph {et~al.}(2015)\citenamefont {White},
  \citenamefont {Scalapino},\ and\ \citenamefont {Kivelson}}]{Kivelson}%
  \BibitemOpen
  \bibfield  {author} {\bibinfo {author} {\bibfnamefont {S.~R.}\ \bibnamefont
  {White}}, \bibinfo {author} {\bibfnamefont {D.~J.}\ \bibnamefont
  {Scalapino}}, \ and\ \bibinfo {author} {\bibfnamefont {S.~A.}\ \bibnamefont
  {Kivelson}},\ }\href {\doibase 10.1103/PhysRevLett.115.056401} {\bibfield
  {journal} {\bibinfo  {journal} {Phys. Rev. Lett.}\ }\textbf {\bibinfo
  {volume} {115}},\ \bibinfo {pages} {56401} (\bibinfo {year}
  {2015})}\BibitemShut {NoStop}%
\end{thebibliography}
%

\onecolumngrid
\appendix

\section{The exact sign structures of the one-hole ground states of the 1D $\sigma$$\cdot$$t$-$J$ and $t$-$J$ models under open boundary condition}
\label{Appen:sign}

In this appendix, we demonstrate that the one-hole ground states of the 1D $\sigma$$\cdot$$t$-$J$ and $t$-$J$ models in the sector $S^z_\mathrm{tot}=-\sigma /2$ satisfy the sign structures given in Eqs.~(\ref{eq:stJ_GSsign}) and (\ref{eq:tJ_GSsign}), respectively.

First of all, the off-diagonal elements of the $\sigma$$\cdot$$t$-$J$ model in the basis $\{(-\sigma)^i c_{i\sigma} |\{s\}\rangle\}$ are all non-positive as shown below:
\begin{align}
  \label{eq:stJ_sign1}\nonumber
  \langle \{s\} | c_{j\sigma}^\dagger (-\sigma)^j (-\sigma t c_{i\sigma}^\dagger c_{j\sigma}) (-\sigma)^i c_{i\sigma} |\{s\}\rangle &= -(-\sigma)^{i+j+1} t \langle \{s\} | n_{i\sigma} n_{j\sigma} |\{s\}\rangle \\\nonumber
  &= - t \langle \{s\} | n_{i\sigma} n_{j\sigma} |\{s\}\rangle\\
  &\le 0,\\
  \label{eq:stJ_sign2}\nonumber
  \langle \{s\} | c_{j\sigma}^\dagger (-\sigma)^j (-\bar\sigma t c_{i\bar\sigma}^\dagger c_{j\bar\sigma}) (-\sigma)^i c_{i\sigma} |\{s'\}\rangle &= (-\sigma)^{i+j+1} t \langle \{s\} | c_{i\bar\sigma}^\dagger c_{i\sigma} c_{j\sigma}^\dagger c_{j\bar\sigma} |\{s'\}\rangle \\\nonumber
  &= t \langle \{s\} | S_i^{\mp} S_j^{\pm} |\{s'\}\rangle\\
  &\le 0,\\
  \label{eq:stJ_sign3}
  \langle \{s\} | c_{h\sigma}^\dagger (-\sigma)^h \left(\frac{J}{2} S_i^+ S_j^-\right) (-\sigma)^h c_{h\sigma} |\{s'\}\rangle &= \frac{J}{2} \langle \{s\} | c_{h\sigma}^\dagger S_i^+ S_j^- c_{h\sigma} |\{s'\}\rangle
  \le 0,
\end{align}
where $i$ and $j$ belong to the nearest neighbors in the $\sigma$$\cdot$$t$-$J$ model and the following property of the Marshall basis $\{|\{s\}\rangle\}$ is used:
\begin{align}
  \label{}
  \langle \{s\} | S_i^{\mp} S_j^{\pm} |\{s'\}\rangle \le 0.
\end{align}

Similarly, the off-diagonal elements of the $t$-$J$ model in the basis $\{ \E^{-\I\hat{\Omega}_i}(-\sigma)^i c_{i\sigma} |\{s\}\rangle\}$ can be also shown to be non-positive. Here, $  \E^{-\I\hat{\Omega}_i}=(-1)^{N_{R(i)}^\downarrow} $, where $N_{R(i)}^\downarrow=\sum_{l>i} n_{l\downarrow} $ denotes the number of down-spins on the right-hand-side of site $i$.
Then, a straightforward manipulation gives rise to
\begin{align}
  \label{eq:tJ_sign1}\nonumber
  \langle \{s\} | c_{j\sigma}^\dagger (-\sigma)^j (-1)^{N_{R(j)}^\downarrow} (-t c_{i\sigma}^\dagger c_{j\sigma}) (-1)^{N_{R(i)}^\downarrow} (-\sigma)^i c_{i\sigma} |\{s\}\rangle &= \sigma (-\sigma)^{i+j} t \langle \{s\} | n_{i\sigma} n_{j\sigma} |\{s\}\rangle \\\nonumber
  &= - t \langle \{s\} | n_{i\sigma} n_{j\sigma} |\{s\}\rangle\\
  &\le 0,\\
  \label{eq:tJ_sign2}\nonumber
  \langle \{s\} | c_{j\sigma}^\dagger (-\sigma)^j (-1)^{N_{R(j)}^\downarrow} (-t c_{i\bar\sigma}^\dagger c_{j\bar\sigma}) (-1)^{N_{R(i)}^\downarrow} (-\sigma)^i c_{i\sigma} |\{s'\}\rangle &= (-\sigma) (-\sigma)^{i+j} t \langle \{s\} | c_{i\bar\sigma}^\dagger c_{i\sigma} c_{j\sigma}^\dagger c_{j\bar\sigma} |\{s'\}\rangle \\\nonumber
  &= t \langle \{s\} | S_i^{\mp} S_j^{\pm} |\{s'\}\rangle\\
  &\le 0,\\
  \label{eq:tJ_sign3}
  \langle \{s\} | c_{h\sigma}^\dagger (-\sigma)^h (-1)^{N_{R(h)}^\downarrow} \left(\frac{J}{2} S_i^+ S_j^-\right) (-1)^{N_{R(h)}^\downarrow} (-\sigma)^h c_{h\sigma} |\{s'\}\rangle &= \frac{J}{2} \langle \{s\} | c_{h\sigma}^\dagger S_i^+ S_j^- c_{h\sigma} |\{s'\}\rangle
  \le 0.
\end{align}

According to the Perron-Frobenius theorem, if all the off-diagonal elements of a Hamiltonian in a given basis are non-positive, then the ground state of the Hamiltonian must have a non-negative coefficient in this basis. Therefore, similar to the ground state of the Heisenberg model at half-filling in the Marshall basis $\{|\{s\}\rangle\}$, the one-hole ground states of the 1D $\sigma$$\cdot$$t$-$J$ and the $t$-$J$ models satisfy Eqs.~(\ref{eq:stJ_GSsign}) and (\ref{eq:tJ_GSsign}), respectively, with the non-negative coefficient $a(i,\{s\})$.

Furthermore, the $t_\uparrow$-$t_\downarrow$-$J$ model Eq.~(\ref{Eq:tutdJmodel}) ($t_\uparrow$ is fixed to be positive) has the same sign structure as the $\sigma$$\cdot$$t$-$J$ model ($t$-$J$ model) if $t_\downarrow/t_\uparrow<0$ ($t_\downarrow/t_\uparrow>0$). Namely, its one-hole ground state wave function always satisfies the same form as Eq.~(\ref{eq:stJ_GSsign}) or Eq.~(\ref{eq:tJ_GSsign}), depending on the sign of $t_\downarrow/t_\uparrow $. Of course, the non-negative coefficient $a(i,\{s\}) \ge 0$ will also depend on the ratio $|t_\downarrow/t_\uparrow |$.

\section{Sign structures of the $\sigma$$\cdot$$t$-$J$ model and the $t$-$J$ model under periodic/anti-periodic boundary conditions}
\label{Appen:sign_PBC}

In this appendix, we analyze the sign structures of the $\sigma$$\cdot$$t$-$J$ model and the $t$-$J$ model under periodic/anti-periodic boundary conditions. We will show that the $\sigma$$\cdot$$t$-$J$ model under periodic boundary condition is still sign-free. On the other hand, the $t$-$J$ model under periodic/anti-periodic boundary condition has an interference effect due to the nontrivial sign structure.

\subsection{The $\sigma$$\cdot$$t$-$J$ model}

First let us consider the $\sigma$$\cdot$$t$-$J$ model on a lattice with even $L$:
\begin{align}
  \label{}\nonumber
  H_{\sigma \cdot t\text{-}J} &= -t \sum_{i=1}^{L-1} \sum_\sigma \sigma{({c_{i\sigma}^{\dag}c_{i+1\sigma }+\mathrm{h.c.}})} +  J\sum_{i=1}^{L-1} \left(\mathbf{S}_{i}\cdot \mathbf{S}_{i+1}-\frac{1}{4}n_{i}n_{i+1}\right) \\
  &\quad -\eta t \sum_\sigma \sigma({c_{1\sigma }^{\dag}c_{L\sigma }+\mathrm{h.c.}}) + J\left(\mathbf{S}_{1}\cdot \mathbf{S}_{L}-\frac{1}{4}n_{1}n_{L}\right),
\end{align}
where $\eta=\pm1$ denotes the periodic or anti-periodic boundary conditions. Under a Marshall transformation $U_M = (-1)^{N_A^\downarrow}$, which is diagonal in the Marshall basis $\{|\{s\}\rangle\}$, the hopping term changes according to $c_{i\sigma }^{\dag}c_{i+1\sigma} \rightarrow \sigma c_{i\sigma }^{\dag}c_{i+1\sigma}$, while the superexchange term changes according to $S_i^+ S_{i+1}^- \rightarrow -S_i^+ S_{i+1}^-$. Therefore, the total Hamiltonian is transformed to
\begin{align}
  \label{}\nonumber
  U_M H_{\sigma \cdot t\text{-}J} U_M^\dagger &= -t \sum_{i=1}^{L-1} \sum_\sigma {({c_{i\sigma}^{\dag}c_{i+1\sigma }+\mathrm{h.c.}})} +  J\sum_{i=1}^{L-1} \left(-\frac{1}{2}(S_i^+ S_{i+1}^- + S_i^- S_{i+1}^+) + S_i^z S_{i+1}^z -\frac{1}{4}n_{i}n_{i+1}\right) \\
  &\quad -\eta t \sum_\sigma ({c_{1\sigma }^{\dag}c_{L\sigma }+\mathrm{h.c.}}) + J\left(-\frac{1}{2}(S_1^+ S_{L}^- + S_1^- S_{L}^+) + S_1^z S_{L}^z - \frac{1}{4}n_{1}n_{L}\right).
\end{align}
Under the periodic boundary condition ($\eta=1$), the off-diagonal terms of the Hamiltonian are always non-positive (with the fermion sign in the hopping term considered for the one-hole case). We therefore conclude that the one-hole-doped $\sigma$$\cdot$$t$-$J$ model is sign-problem-free not only under the open boundary condition, but also under the periodic boundary condition. In the partition function language, the partition function can be written as:
\begin{align}
  \label{}
  Z_{\sigma\cdot t\text{-}J}^\mathrm{PBC} = \sum_c \mathcal{Z}[c],
\end{align}
where $\mathcal{Z}[c]$ is non-negative and $c$ is the world-line path of the system.

For the anti-periodic boundary condition ($\eta=-1$), the only non-trivial sign comes from the process that a hole hopping through the boundary. The partition function is
\begin{align}
  \label{}
  Z_{\sigma\cdot t\text{-}J}^\mathrm{ABC} = \sum_c (-1)^{N_\mathrm{bdy}^h} \mathcal{Z}[c],
\end{align}
where $\mathcal{Z}[c]$ is non-negative and $N_\mathrm{bdy}^h$ is the number of times that the hole crosses the boundary.
Therefore, in the squeezed spin chain approximation, the energy difference between the anti-periodic and periodic boundary condition $\Delta E_{G}^{\text{1-hole}}$ is merely the energy difference for a Bloch particle under anti-periodic and periodic boundary conditions. The scaling of the latter is given by
\begin{align}
  \label{}
  \Delta E_{G}^{\text{1-hole}} \propto 1-\cos\left(\frac{\pi}{L}\right) \propto \frac{1}{L^2}.
\end{align}
This $L^{-2}$ scaling of the energy difference is confirmed by the DMRG results in Fig.~\ref{Fig:PBC_E0} (c).

\subsection{The $t$-$J$ model}

Now let us turn to the $t$-$J$ model under the periodic boundary condition:
\begin{align}
  \nonumber
  H_{t\text{-}J} &= -t \sum_{i=1}^{L-1} \sum_\sigma {({c_{i\sigma }^{\dag}c_{i+1\sigma }+\mathrm{h.c.}})} +  J\sum_{i=1}^{L-1} \left(\mathbf{S}_{i}\cdot \mathbf{S}_{i+1}-\frac{1}{4}n_{i}n_{i+1}\right) \\
  &\quad -\eta t \sum_\sigma ({c_{1\sigma }^{\dag}c_{L\sigma }+\mathrm{h.c.}}) + J\left(\mathbf{S}_{1}\cdot \mathbf{S}_{L}-\frac{1}{4}n_{1}n_{L}\right).
\end{align}
The Marshall transformation $U_M = (-1)^{N_A^\downarrow}$ changes the above Hamiltonian to
\begin{align}
  \label{}\nonumber
  U_M H_{t\text{-}J} U_M^\dagger &= -t \sum_{i=1}^{L-1} \sum_\sigma \sigma{({c_{i\sigma}^{\dag}c_{i+1\sigma }+\mathrm{h.c.}})} + J\sum_{i=1}^{L-1} \left(-\frac{1}{2}(S_i^+ S_{i+1}^- + S_i^- S_{i+1}^+) + S_i^z S_{i+1}^z -\frac{1}{4}n_{i}n_{i+1}\right) \\
  &\quad -\eta t \sum_\sigma \sigma({c_{1\sigma }^{\dag}c_{L\sigma }+\mathrm{h.c.}}) + J\left(-\frac{1}{2}(S_1^+ S_{L}^- + S_1^- S_{L}^+) + S_1^z S_{L}^z - \frac{1}{4}n_{1}n_{L}\right).
\end{align}
Now to absorb the sign in the front of the bulk hopping term, i.e. $\sigma c_{i\sigma}^{\dag} c_{i+1\sigma} \rightarrow c_{i\sigma}^{\dag} c_{i+1\sigma}$, we perform the phase string transformation $U_{PS} \equiv \E^{\I\hat{\Theta}} = \prod_{i<l}(-1)^{n_i^h n_{l\downarrow}}$. The Hamiltonian is further transformed to
\begin{align}
  \label{}\nonumber
  U_{PS} U_M H_{t\text{-}J} U_M^\dagger U_{PS}^\dagger &= -t \sum_{i=1}^{L-1} \sum_\sigma {({c_{i\sigma}^{\dag}c_{i+1\sigma}+\mathrm{h.c.}})} + J\sum_{i=1}^{L-1} \left(-\frac{1}{2}(S_i^+ S_{i+1}^- + S_i^- S_{i+1}^+) + S_i^z S_{i+1}^z -\frac{1}{4}n_{i}n_{i+1}\right) \\
  &\quad -\eta t \sum_\sigma \sigma (-1)^{N_\downarrow}({c_{1\sigma }^{\dag}c_{L\sigma}+\mathrm{h.c.}}) + J\left(\frac{1}{2}(S_1^+ S_{L}^- + S_1^- S_{L}^+) + S_1^z S_{L}^z - \frac{1}{4}n_{1}n_{L}\right).
\end{align}
Note that the boundary hopping term $c_{1\sigma }^{\dag}c_{L\sigma}$ ($\sigma=\pm$) and the boundary superexchange term $S_1^+ S_{L}^-$ acquire a additional sign $(-1)^{N_\downarrow}$ and $-1$, respectively. Here $(-1)^{N_\downarrow}$ denotes the total number of down spins.

From the analysis above, under the periodic boundary condition ($\eta=1$), the non-trivial sign is $t c_{1\downarrow}^{\dag}c_{L\downarrow}$ ($t c_{1\uparrow}^{\dag}c_{L\uparrow}$) if $N_\downarrow$ is even (odd) and $(J/2) (S_1^+ S_{L}^- + S_1^- S_{L}^+)$. The partition function is given by
\begin{align}
  \label{eq:Z_tJ_PBC}
  Z_{t\text{-}J} = \sum_c (-1)^{N_\mathrm{bdy}^{\uparrow/\downarrow}} \mathcal{Z}[c],
\end{align}
where $\mathcal{Z}[c]$ is non-negative and $N_\mathrm{bdy}^{\uparrow/\downarrow}$ (for $N_\downarrow$ odd/even) is the number of times that a up/down spin crosses the boundary.
On the other hand, under the anti-periodic boundary condition ($\eta=-1$), the partition function is still Eq.~(\ref{eq:Z_tJ_PBC}). But $N_\mathrm{bdy}^{\uparrow}$ ($N_\mathrm{bdy}^{\downarrow}$) is for $N_\downarrow$ even (odd).

Since the number of down spins increases by one, if one increases the chain length $L$ by two, $N_\mathrm{bdy}^{\uparrow}$ and $N_\mathrm{bdy}^{\downarrow}$ in the partition function are switched. Effectively, the periodic and anti-periodic boundary conditions are also switched (the combined quantity $\eta (-1)^{N_\downarrow}$ determines the sign structure). Therefore, one expects the energy difference between the anti-periodic and periodic boundary condition $\Delta E_{G}^{\text{1-hole}}$ to oscillate with increasing $L$. The DMRG result indeed shows the oscillation and $L^{-3}$ scaling of the energy difference $\Delta E_{G}^{\text{1-hole}}$ in Fig.~\ref{Fig:PBC_E0}(b).

\section{Sign structure of the variational wave functions in Eq.~(\ref{eq:wf1})}
\label{Appen:sign_wf}

In this appendix, we will show explicitly that the variational wave function Eq.~(\ref{eq:wf1}) satisfies the sign structure requirement of the $t$-$J$ model. The hole wave function $\varphi_h(i)$ will be also determined by minimizing the hopping energy.

Since the spin and charge are totally separated in the squeezed spin chain approximation, the state obtained from moving the hole from site $i$ to $j$ is exactly the one with hole at $j$ initially. To be more precise, the matrix element at each step of hopping energy for the $t$-$J$ model under Eq.~(\ref{eq:wf1}), is given by
\begin{align}
  \label{eq:tJ_t}
  \langle\Psi| (-t c_{i\sigma}^\dagger c_{j\sigma}) |\Psi\rangle_{t\text{-}J}
  = \varphi_h(j)\varphi_h(i) \left(\langle\phi_0| \hat T_j^\dagger \right) (-t c_{i\sigma}^\dagger c_{j\sigma}) \left( \hat T_i |\phi_0\rangle\right)
  = \varphi_h(j)\varphi_h(i) (-t/2) \le 0,
\end{align}
where $1/2$ comes from the two possibilities of the spin on sites $i$ and $j$. As a result, to minimize the hopping energy, we choose $\varphi_h(i)$ ($i=1,2,\cdots,L$) to be the Bloch state for a free particle in the tight binding model:
\begin{align}
  \label{eq:phi}
  \varphi_h(i) =
  \begin{cases}
  \sqrt\frac{1}{L}, & \text{for periodic boundary condition}, \\
  \sqrt{\frac{2}{L+1}}\sin\left(\frac{i\pi}{L+1}\right), & \text{for open boundary condition}.
  \end{cases}
\end{align}

As for the superexchange terms of the $t$-$J$ model, the off-diagonal terms, such as $S_i^+ S_j^-$, always change the number of down spins on $A$ sublattice by one, while leave the relative positions of the hole and spins unchanged. The Marshall sign mismatch in this process in the one-hole-doped case is exactly the same as the half-filled spin chain wave function. Therefore, the off-diagonal terms of the superexchange terms are also non-positive for the hole wave function basis in the variational wave function $|\Psi\rangle_{t\text{-}J}$, which inherits the Marshall sign from the half-filled ground state $|\phi_0\rangle$.

Thus, similar to the basis states used in Appendix~\ref{Appen:sign}, the single-hole wave function basis $\{ \hat T_i |\phi_0\rangle \}$ in the variational wave function $|\Psi\rangle_{t\text{-}J}$ in Eq.~(\ref{eq:wf1}) is indeed the sign-free basis of the $t$-$J$ model. Namely the variational wave function $|\Psi\rangle_{t\text{-}J}$ with non-negative hole wave function $\varphi_h(i)$ satisfies the sign structure requirement of the model. The same argument can be apply to the $\sigma$$\cdot$$t$-$J$ model as its one-hole ground state $|\Phi\rangle_{\sigma\cdot t\text{-}J}$ is connected to  $|\Psi\rangle_{t\text{-}J}$ by a unitary transformation Eq.~(\ref{eq:wf12}).

\section{Monte Carlo for variational wave functions}
\label{Appen:MC}

In this appendix, we present the Monte Carlo method in calculating $n_{k\alpha}$ for the variational wave functions $|\Psi\rangle_{t\text{-}J}$ and $|\Phi\rangle_{\sigma{\cdot}t\text{-}J}$. The procedures have some similarities to those used in Ref.~\onlinecite{ladder2015}.

The single-hole-doped basis is constructed from the singlet valence bond state $|v\rangle=\sum_{\{\sigma\}} \left(\prod_{(ij)\in v} \epsilon_{\sigma_i,\sigma_j}\right) c_{1\sigma_1}^\dagger ... c_{L\sigma_L}^\dagger |0\rangle$ specified by the dimer covering configuration $v$. By acting the operator $\hat T_h$, the spins on site $x\ge h$ are translated by one lattice constant along $\hat x$ direction, effectively creating a hole at site $h$. We denote this single-hole-doped valence bond state by $|h,v\rangle$, which is defined on a lattice with $L$ sites ($L$ is odd because of adding a hole site). The half-filled Liang-Doucot-Anderson type RVB state can be obtained by Monte Carlo as \cite{LDA1988,Sandvik2010}
\begin{align}
  \label{eq:phi0}
  |\phi_0\rangle = \sum_v w_v |v\rangle,
\end{align}
where $w_v = \prod_{(ij)\in v} h_{ij}$, with non-negative variational parameters $h_{ij}$'s, is the non-negative coefficient associated with the valence bond state $|v\rangle$. The variational wave function $|\Psi\rangle_{t\text{-}J}$ for the squeezed $t$-$J$ chain is then given by
\begin{align}
  \label{}
  |\Psi\rangle_{t\text{-}J} = \sum_h \varphi_h(h) \sum_v w_v |h,v\rangle,
\end{align}
where $w_v$ is the same as in Eq.~(\ref{eq:phi0}), and the normalized hole wave function $\varphi_h(h)$ is chosen according to Eq.~(\ref{eq:phi}). Therefore there are in fact no variational parameters to tune in $|\Psi\rangle_{t\text{-}J}$.

The normalization of the basis states is given by
\begin{align}
  \label{}
  \langle h',v'|h,v\rangle = \delta_{hh'} \langle v'|v\rangle = 2^{N_{v,v'}^\mathrm{HF}},
\end{align}
where $N_{v,v'}^\mathrm{HF}$ is the number of loops in the transition graph of the dimer covers $v$ and $v'$ at half-filling. The normalization of the wave function $|\Psi\rangle_{t\text{-}J}$ is then
\begin{align}
  \label{}
  \langle \Psi|\Psi\rangle_{t\text{-}J} = \sum_h \varphi_h(h)^2 \sum_{v,v'} \langle h',v'|h,v\rangle = \left( \sum_h \varphi_h(h)^2 \right) \sum_{v,v'} w_{v'} w_v 2^{N_{v,v'}^\mathrm{HF}} = \sum_{v,v'} w_{v'} w_v 2^{N_{v,v'}^\mathrm{HF}},
\end{align}
where we used the normalization of the hole wave function $\varphi_h(h)$.

From the above formula for the normalization of $|\Psi\rangle_{t\text{-}J}$, we can use MC method to calculate the momentum distribution of $|\Psi\rangle_{t\text{-}J}$:
\begin{align}
  \label{eq:nk}
  \langle n_{k\alpha}\rangle = \langle c_{k\alpha}^\dagger c_{k\alpha} \rangle = \frac{1}{L} \sum_{i,j} \E^{\mathrm i k(i-j)} \langle c_{i\alpha}^\dagger c_{j\alpha} \rangle.
\end{align}
We should first calculate
\begin{align}
  \label{}\nonumber
  \frac{\langle \Psi| c_{i\alpha}^\dagger c_{j\alpha} |\Psi\rangle_{t\text{-}J}}{\langle \Psi|\Psi\rangle_{t\text{-}J}}
  &= \frac{ \sum_{h,h'}\varphi_h(h')\varphi_h(h) \sum_{v,v'} w_{v'}w_v \langle h',v'| c_{i\alpha}^\dagger c_{j\alpha} |h,v\rangle }{ \sum_{v,v'} w_{v'} w_v 2^{N_{v,v'}^\mathrm{HF}} } \\
  &= \frac{ \sum_{v,v'} w_{v'} w_v 2^{N_{v,v'}^\mathrm{HF}} T_{ij\alpha}(v,v') }{ \sum_{v,v'} w_{v'} w_v 2^{N_{v,v'}^\mathrm{HF}} },
\end{align}
where we have defined
\begin{align}
  \label{eq:Tijs}
  T_{ij\alpha}(v,v') \equiv \sum_{h,h'} \varphi_h(h')\varphi_h(h) \frac{1}{2^{N_{v,v'}^\mathrm{HF}}} \langle h',v'| c_{i\alpha}^\dagger c_{j\alpha} |h,v\rangle.
\end{align}
The positive value $w_{v'} w_v 2^{N_{v,v'}^\mathrm{HF}}$ can be used as the distribution weight in the MC procedure. The quantity $T_{ij\alpha}(v,v')$ is averaged over to obtain $\langle c_{i\alpha}^\dagger c_{j\alpha} \rangle$, and then the momentum distribution $\langle n_{k\alpha}\rangle$ according to Eq.~(\ref{eq:nk}).

Now our task is to simplify $T_{ij\alpha}(v,v')$ in Eq.~(\ref{eq:Tijs}) to be calculated numerically. For $i=j$, we have
\begin{align}
  \label{eq:Tiis}
  T_{ii\alpha}(v,v') = \sum_{h,h'} \varphi_h(h')\varphi_h(h) \frac{1}{2^{N_{v,v'}^\mathrm{HF}}} \delta_{hh'} \delta_{i\neq h} 2^{N_{v,v'}^\mathrm{HF}-1}
  = \sum_{h\neq i} \frac{1}{2} \varphi_h(h)^2 = \frac{1}{2} \left( 1-\varphi_h(i)^2 \right).
\end{align}
This is simply the probability $1-\varphi_h(i)^2$ of finding an electron at site $i$ divided by two, because of two spin values ($\alpha=\pm$). On the other hand, for $i\neq j$, we have
\begin{align}
  \label{}\nonumber
  \langle h',v'| c_{i\alpha}^\dagger c_{j\alpha} |h,v\rangle &= - \delta_{ih} \delta_{jh'} (\langle j,v'| c_{j\alpha}) (c_{i\alpha}^\dagger |i,v\rangle)\\\nonumber
  &= - \delta_{ih} \delta_{jh'} \sum_{\{s\},s_i=s_j=\alpha} \delta_{v,\{s\}}\delta_{v',\{s\}} (-1)^{i+j} \eta(\{s\}_i)\eta(\{s\}_j)\\
  &= - \delta_{ih} \delta_{jh'} \sum_{\{s\},s_i=s_j=\alpha} \delta_{v,\{s\}}\delta_{v',\{s\}} (-1)^{i+j} (-1)^{N_{(i,j)}^\downarrow} \alpha^{i+j-1}.
\end{align}
Note that in the first line, $c_{i\alpha}^\dagger |i,v\rangle$ is a half-filled spin state with fermion sign $(-1)^{i-1}$ (coming from moving $c_{i\alpha}^\dagger$ to the $i$-th place in the sequence $c_{1s_1}^\dagger \cdots c_{i-1s_{i-1}}^\dagger c_{i+1s_{i+1}}^\dagger \cdots c_{Ls_L}^\dagger$) and the Marshall sign $\eta(\{s\}_i)$ on the original lattice with $L-1$ sites (removing the $i$ site). $\delta_{v,\{s\}}$ specifies the constraint that the summation over the spin configuration $\{s\}$ should be compatible with the dimer cover $v$, i.e., the spins on sites belonging to the same dimer should be opposite. From the second line to the third line, we simplified the product of the Marshall signs $\eta(\{s\}_i)\eta(\{s\}_j)$ for the initial and final valence bond states $|v\rangle$ and $|v'\rangle$: The Marshall signs of the two states for spins on site $x<i$ or $x>j$ (suppose $i<j$) cancel each other; The Marshall signs for sites $i$ and $j$ contribute the factor $\alpha^{i+j-1}$; The product of Marshall signs for sites $i<x<j$ is $(-1)^{N_{(i,j)}^\downarrow}$, where $N_{(i,j)}^\downarrow$ is the number of down spins on sites $i<x<j$ for the spin configuration $\{s\}$. As a result, Eq.~(\ref{eq:Tijs}) can be reduced to
\begin{align}
  \label{eq:Tijs1}
  T_{ij\alpha}(v,v') = \varphi_h(j)\varphi_h(i) \frac{1}{2^{N_{v,v'}^\mathrm{HF}}} \sum_{\{s\},s_i=s_j=\alpha} \delta_{v,\{s\}}\delta_{v',\{s\}} (-1)^{i+j+1} (-1)^{N_{(i,j)}^\downarrow} \alpha^{i+j-1}
\end{align}
for $i\neq j$.

For the variational wave function $|\Phi\rangle_{\sigma\cdot t\text{-}J}$ of the $\sigma$$\cdot$$t$-$J$ model, the explicit form in the single-hole-doped valence bond basis reads
\begin{align}
  \label{}
  |\Phi\rangle_{\sigma\cdot t\text{-}J} = \sum_h \varphi_h(h) \sum_v (-1)^{N_{R(h)}^\downarrow} w_v |h,v\rangle,
\end{align}
where $(-1)^{N_{R(h)}^\downarrow}$ comes from the phase string transformation connecting the ground states of the $t$-$J$ model and the $\sigma$$\cdot$$t$-$J$ model. Following the same analysis above, we can also use the positive value $w_{v'} w_v 2^{N_{v,v'}^\mathrm{HF}}$ as the distribution weight in the MC simulations, and calculate the average value of the quantity $T_{ij\alpha}(v,v')$. For $i=j$, the formula Eq.~(\ref{eq:Tiis}) is the same. For $i\neq j$, however, there is an additional phase string factor $(-1)^{N_{(i,j)}^\downarrow}$ from sites $i<x<j$ and $\sigma$ from the site $j$. The final result is therefore
\begin{align}
  \label{eq:Tijs2}
  T_{ij\alpha}(v,v') = \varphi_h(j)\varphi_h(i) \frac{1}{2^{N_{v,v'}^\mathrm{HF}}} \sum_{\{s\},s_i=s_j=\alpha} \delta_{v,\{s\}}\delta_{v',\{s\}} (-1)^{i+j+1} \alpha^{i+j}.
\end{align}

In summary, to calculate the momentum distribution $n_{k\alpha}$ of the wave functions $|\Psi\rangle_{t\text{-}J}$ and $|\Phi\rangle_{\sigma\cdot t\text{-}J}$, we first simulate a distribution of valence bond states with (unnormalized) probability $w_{v'} w_v 2^{N_{v,v'}^\mathrm{HF}}$ by MC. This procedure is similar to the half-filled case \cite{LDA1988,Sandvik2010}. Then $T_{ij\alpha}(v,v')$ is averaged over this distribution by using Eqs.~(\ref{eq:Tiis}), (\ref{eq:Tijs1}) and (\ref{eq:Tijs2}). The momentum distribution is finally calculated following Eq.~(\ref{eq:nk}).

\end{document}